\documentclass[fleqn,usenatbib]{mnras}

\usepackage{newtxtext,newtxmath}

\usepackage[T1]{fontenc}
\usepackage[utf8]{inputenc}
\DeclareRobustCommand{\VAN}[3]{#2}
\let\VANthebibliography\thebibliography
\def\thebibliography{\DeclareRobustCommand{\VAN}[3]{##3}\VANthebibliography}

\usepackage{graphicx}	
\usepackage{amsmath}	
\usepackage{subcaption}  
\usepackage{float}
\usepackage{natbib}

\title[Baryonic effects on the WL scattering transform]{FLAMINGO: Baryonic effects on the weak lensing scattering transform}

\author[M. Marinichenko et al.]{
Mariia Marinichenko,$^{1}$\thanks{E-mail: marinichenko@strw.leidenuniv.nl (MM)}
Marcel P. van Daalen,$^{1}$
Elena Sellentin,$^{1,2}$
Jeger C. Broxterman,$^{1,3}$
\and Matthieu Schaller,$^{1,3}$
Joop Schaye$^{1}$
\\
$^{1}$ Leiden Observatory, Leiden University, PO Box 9513, NL-2300 RA Leiden, the Netherlands\\
$^{2}$ Mathematical Institute, Leiden University, PO Box 9512, NL-2300 RA Leiden, the Netherlands\\
$^{3}$ Lorentz Institute for Theoretical Physics, Leiden University, PO Box 9506, NL-2300 RA Leiden, the Netherlands\\
}

\date{Accepted XXX. Received YYY; in original form ZZZ}

\pubyear{\the\year{}}

\begin{document}
\label{firstpage}
\pagerange{\pageref{firstpage}--\pageref{lastpage}}
\maketitle
\begin{abstract}
The scattering transform is a wavelet-based statistic capable of capturing non-Gaussian features in weak lensing (WL) convergence maps and has been proven to tighten cosmological parameter constraints by accessing information beyond two-point functions. However, its application in cosmological inference requires a clear understanding of its sensitivity to astrophysical systematics, the most significant of which are baryonic effects. These processes substantially modify the matter distribution on small to intermediate scales ($k\gtrsim 0.1\,h\,\mathrm{Mpc}^{-1}$), leaving scale-dependent imprints on the WL convergence field.
We systematically examine the impact of baryonic feedback on scattering coefficients using full-sky WL convergence maps with Stage IV survey characteristics, generated from the FLAMINGO simulation suite. These simulations include a broad range of feedback models, calibrated to match the observed cluster gas fraction and galaxy stellar mass function, including systematically shifted variations, and incorporating either thermal or jet-mode AGN feedback. We characterise baryonic effects using a baryonic transfer function defined as the ratio of hydrodynamical to dark-matter-only scattering coefficients. While the coefficients themselves are sensitive to both cosmology and feedback, the transfer function remains largely insensitive to cosmology and shows a strong response to feedback, with suppression reaching up to $10\%$ on scales of $k\gtrsim 0.1\,h\,\mathrm{Mpc}^{-1}$. 
We also demonstrate that shape noise significantly diminishes the sensitivity of the scattering coefficients to baryonic effects, reducing the suppression from $\sim 2 - 10 \;\%$ to $\sim 1\;\%$, even with 1.5 arcmin Gaussian smoothing. This highlights the need for noise mitigation strategies and high-resolution data in future WL surveys. 
\end{abstract}



\section{Introduction}

The large-scale structure (LSS) is an ample source of information about the content and expansion history of the Universe. A key observable for accessing this information is cosmic shear, subtle distortions in the shapes of distant galaxies caused by weak gravitational lensing \citep[WL; for reviews, see][]{bartelmann_weak_2001,2008ARNPS..58...99H,kilbinger_cosmology_2015}.

Since WL directly traces the total matter distribution, it serves as a powerful probe of cosmology. While the lensing field remains nearly Gaussian on sufficiently large scales of comoving wavenumbers $k \lesssim 0.1 \: h\: \text{Mpc}^{-1}$, the non-linear formation of cosmic structures induces non-Gaussian features at smaller scales. Moreover, the structure formation is also strongly affected by astrophysical processes within galaxies, which impact matter clustering~\citep[e.g.,][]{2011MNRAS.415.3649V, 2020MNRAS.491.2424V, Chisari_2019, 2023MNRAS.523.2247S}. 

With the improved resolution of modern WL surveys, the small scales, $k\gtrsim 0.1\,h\,\mathrm{Mpc}^{-1}$, are now accessible with high precision. 
Accurate measurement of their features is essential for recovering the total cosmological information and requires advanced statistical tools beyond standard two-point functions. To address this issue, a wide range of higher-order statistics (HOS), often referred to as non-Gaussian statistics, have been developed. A recent parameter inference from non-Gaussian statistics was presented in \cite{2025JCAP...01..006C}, using the year-1 data of the Hyper Suprime Cam \citep[HSC;][]{2018PASJ...70S...4A}. Other surveys which lend themselves to such analyses are the Kilo Degree Survey \citep[KiDS; ][]{2015MNRAS.454.3500K},  the Dark Energy Survey \citep[DES;][]{2016MNRAS.460.1270D}, with the long-term goal being an inference of cosmological parameters from non-Gaussianity estimators of Stage IV surveys such as Euclid~\citep[]{2023A&A...675A.120E}, the Roman Space Telescope \citep{2015arXiv150303757S}, and the Rubin Observatory survey \citep[LSST;][] {2009arXiv0912.0201L}.

Examples of such non-Gaussianity estimators are Minkowski functionals \citep[e.g.,][]{2019JCAP...06..019M,2020A&A...633A..71P, Grewal_2022}, higher-order moments \citep[e.g.,][]{PhysRevD.98.023507, PhysRevD.98.023508, 2022PhRvD.106h3509G}, peaks \citep[e.g.,][]{PhysRevD.102.103531, PhysRevD.99.063527, PhysRevD.94.043533, 2024MNRAS.534.3305H, 2024MNRAS.528.4513M}, the one-point probability distribution function (PDF) \citep[e.g.,][]{PhysRevD.102.123545, 2023MNRAS.520.1721G, Castiblanco_2024, 2024JCAP...03..060B}, scattering transform (ST) coefficients \citep[e.g.,][]{2020MNRAS.499.5902C, 2020PhRvD.102j3506A, 2021MNRAS.507.1012C, 2022PhRvD.106j3509V, 2025JCAP...01..006C}, and topological descriptors like Betti numbers \citep[e.g.,][]{2019JCAP...09..052F}. 

These statistics have the potential to substantially tighten constraints on cosmological parameters \citep{ 2023A&A...675A.120E} 
and contributed to distinguishing cosmological effects from astrophysical systematics \citep[e.g.,][]{2013MNRAS.434..148S}.  

Although non-Gaussian estimators have demonstrated strong performance, none of them has proven consistently optimal. Due to the inherent complexity of non-Gaussian fields, each method responds differently to features of the cosmological signal and observational systematics. This diversity has motivated extensive comparisons in recent WL surveys \citep[e.g.,][]{2023A&A...675A.120E, 2025PhRvD.111f3504G}, indicating compromises between robustness to noise and outliers, interpretability of the extracted features, and completeness in capturing the underlying cosmological information.

All of these non-Gaussianity estimators have in common that baryonic feedback impacts the inference of parameters, with the Stage IV adopted removal of feedback being presented in \cite{2025MNRAS.536.2064G}. Such accounting for baryonic feedback profits from the establishment of a baryonic feedback transfer function, which is the goal of this paper.

In this paper, we focus on the scattering transform (ST), a combination of wavelet analysis and convolutional neural networks (CNNs) \citep{2020MNRAS.499.5902C}. Initially introduced by \cite{2012arXiv1203.1513B}, the ST has recently been adapted to cosmology and astrophysics \citep[e.g.,][]{2021arXiv211201288C}, establishing itself as a promising tool for analysing WL data from upcoming surveys. Unlike CNNs, it does not require training and produces a structured set of multi-scale coefficients that capture non-Gaussian features of the input field. These coefficients are both stable under small geometric deformations and robust to noise and outliers, making the ST particularly suited to characterising the complex structures in WL maps \citep[e.g.,][]{2021MNRAS.507.1012C, 2025JCAP...01..006C}. Moreover, the ST offers direct interpretability and incorporates both cross-scale correlations and directional information, providing key advantages over commonly used HOS such as peak counts and topological descriptors. The ST approach is further strengthened by the development of related methods, such as phase harmonics \citep[e.g.,][]{2019A&A...629A.115A, 2024A&A...681A...1A}, scattering spectra~\citep[e.g.,][]{2023arXiv230617210C}, and wavelet moments \citep[e.g.,][]{2022arXiv220407646E}, which have been applied to cosmological simulations and demonstrated the high performance of this HOS family.

However, the sensitivity of the ST to non-Gaussian features is deeply intertwined with contamination by small-scale systematics. The dominant astrophysical systematic is baryonic feedback, a collective term for astrophysical processes such as supernova-driven outflows, active galactic nuclei (AGN) feedback, gas cooling, and star formation. These mechanisms redistribute matter within halos, leaving an imprint in the WL field. For instance, AGN feedback typically reduces the matter power spectrum at intermediate scales ($k \sim 0.1 - 10\: h\: \text{Mpc}^{-1}$), while cooling-driven processes may increase it at smaller scales \citep[e.g.,][]{2011MNRAS.415.3649V, Chisari_2019, 2020MNRAS.491.2424V, PhysRevD.107.023514, 2025MNRAS.539.1337S}. 
Due to its composite nature and dependence on loosely constrained subgrid models, baryonic feedback remains challenging to model analytically and constitutes a major source of uncertainty for non-Gaussian WL analyses, biasing the cosmological inference \citep[e.g., ][]{2013MNRAS.434..148S, 2020MNRAS.495.2531C, 2021A&A...648A.115M, 2022MNRAS.509.3868H, 2024PhRvD.110j3539G, 2025MNRAS.536.2064G}. These effects are considered to be one of the contributors to the tension in the $S_8$ parameter between WL and the cosmic microwave background (CMB) measurements~\citep[e.g.,][]{2024PhRvD.110j3539G}.

Previous generations of WL surveys addressed this problem by applying conservative angular scale cuts, effectively removing small-scale modes \citep[e.g.,][]{ 2018PhRvD..98d3526A, 2022PhRvD.105b3520A, PhysRevD.108.123519}. 
However, such strategies significantly compromise the efficiency of inference, especially in Stage IV surveys where the statistical precision relies on non-linear scales. To retain the information, two main strategies have been proposed. One is based on full hydrodynamical simulations that model baryonic processes using sub-grid prescriptions calibrated on observational data. This approach remains the most comprehensive way to characterise baryonic uncertainty due to its physical consistency and built-in cross-checks with other probes (e.g., X-rays and the Sunyaev–Zeldovich effect). However, it is resource-intensive, and different simulation suites vary in resolution, box size, and calibration targets, producing a wide range of predictions for the amplitude and scale dependence of feedback. A faster approach is the baryonic correction model (BCM), which adjusts dark matter only (DMO) simulations to imitate baryonic effects \citep[e.g.,][]{2020JCAP...04..019S, 2020MNRAS.495.4800A, 2025arXiv250707892S}. 
Although more efficient, it has shown limitations when applied to higher-order statistics such as peak counts~\citep{2023MNRAS.519..573L}. Nonetheless, with calibration based on HOS, they remain a promising approach \citep{2021MNRAS.506.3406L}.

While the impact of baryonic feedback on two-point statistics has been extensively studied, its effect on higher-order statistics remains an open field for exploration~\citep[e.g.,][]{2013MNRAS.434..148S, 2021A&A...648A.115M, 2022MNRAS.509.3868H, 2023A&A...671A..17A, 2025arXiv250507949Z, 2025arXiv250618974B}. Only a few phenomenological approaches, such as those based on halo models \citep[e.g.,][]{PhysRevD.105.023505, 2023OJAp....6E..39A} or symbolic parametrization \citep[e.g.,][]{2025arXiv250608783K}, have been proposed. Several works have examined peak counts in hydrodynamical simulations, revealing strong sensitivity to baryonic processes and degeneracies with other effects, such as massive neutrinos~\citep[e.g.,][]{PhysRevD.84.043529, 2019MNRAS.488.3340F, 2020MNRAS.495.2531C, 2023MNRAS.524.5591F, 2024MNRAS.529.2309B, 2025MNRAS.538..755B}. The sensitivity of the ST to baryonic feedback has received even less attention, which is unsurprising given the relative novelty of the method. 

In this paper, we systematically study the influence of baryonic feedback on ST coefficients using the FLAMINGO simulation suite \citep{2023MNRAS.526.4978S, 2023MNRAS.526.6103K}. This suite contains a range of physically motivated feedback models, calibrated with machine learning to match observed cluster gas fractions and galaxy stellar mass function. We compute the ST coefficients from full-sky convergence maps, generated through backward ray-tracing by \cite{2024MNRAS.529.2309B}, which replicates the conditions of Stage IV WL surveys. By analysing these maps across different feedback scenarios and cosmological models, we assess the scale-dependent imprint of baryonic physics and produce correction factors, which are defined as the ratios of average ST coefficients from hydrodynamical maps to those from their dark matter-only counterparts. This analysis provides new insights into the robustness of the ST and its potential for application in next-generation WL studies.

This paper is organised as follows. In Section~\ref{sec:simulation}, we describe the FLAMINGO simulation suite and the construction of the WL convergence maps used in our analysis. Section~\ref{sec:scattering_transform} introduces the scattering transform, with the details of its implementation. In Section~\ref{sec:results} we present our main results on the impact of baryonic processes on the scattering coefficients across different feedback models and cosmologies. Section~\ref{sec:summary} provides the summary of our key findings and their implications for future surveys.

\section{FLAMINGO simulations and WL convergence maps}

\label{sec:simulation}

The transfer functions presented in this work are computed from WL convergence maps generated using the FLAMINGO simulation suite, which is a set of large-volume cosmological simulations that systematically vary resolution, cosmology, and baryonic feedback models. Below we briefly highlight the main features of the simulations, while a detailed overview is provided in \cite{2023MNRAS.526.4978S}.

The FLAMINGO simulations were performed with the Swift cosmological hydrodynamics code \citep{2024MNRAS.530.2378S} using the \texttt{SPHENIX} scheme for smoothed particle hydrodynamics \citep[SPH;][]{2022MNRAS.511.2367B}. The simulations include massive neutrinos treated with the $\delta f$ method developed by \cite{2021MNRAS.507.2614E}. Initial conditions are generated at $z=31$ with initial conditions generated using \texttt{MonofonIC} \citep{2021MNRAS.503..426H, Michaux_2020}, applying third-order Lagrangian perturbation theory to evolve cold dark matter, baryons, and neutrinos as distinct fluids \citep{2022MNRAS.516.3821E}. 

The simulations exploits subgrid models of baryonic processes that are unresolved at the grid scale. These include radiative cooling and heating per element \citep{2020MNRAS.497.4857P}, star formation compensated for gas pressure \citep{2008MNRAS.383.1210S}, and stellar mass loss over time \citep{2009MNRAS.399..574W}. Stellar and supernova (SN) feedback are applied in kinetic form by transferring momentum to nearby particles, conserving both linear and angular momentum \citep{Dalla_Vecchia_2008, Chaikin_2023}. Supermassive black hole (BH) growth and AGN feedback are modelled following the prescriptions from \cite{springel_simulations_2005} and \cite{ 2009MNRAS.398...53B} with recent extensions by \citet{Bah__2022}. Most runs use thermal, isotropic AGN feedback. However, two simulations apply the kinetic jet mode, where energy is injected by kicking gas particles in the direction of the BH spin with a fixed jet velocity \citep{Hu_ko_2022}.

Calibration of the baryonic physics was performed to reproduce the observed galaxy stellar mass function (SMF) at $z = 0$ and gas fractions in low-redshift galaxy clusters inferred from pre-eROSITA X-ray surveys~\citep[table 5 of ][]{2023MNRAS.526.6103K} and WL observations~\citep[][]{2022PASJ...74..175A}. To achieve this, the efficiency and scale of feedback was controlled by four adjustable subgrid parameters: the SN energy fraction transferred to the ISM ($f_{\rm SN}$), the target wind velocity for SN feedback ($\Delta v_{\rm SN}$), the boost factor for BH accretion rates ($\beta_{\rm BH}$), and the AGN heating temperature ($\Delta T_{\rm AGN}$) or, in the case of jet feedback, the jet velocity ($v_{\rm jet}$).  Rather than tuning parameters manually, the calibration was done through a Gaussian Process emulator trained on a Latin hypercube of small-volume simulations \citep{2023MNRAS.526.6103K}. In addition, this approach allows interpolation in the subgrid parameter space and accounts for observational errors and biases, producing models with explicitly controlled SMF
and cluster gas fractions.

\subsection{Variations}

The FLAMINGO suite provides a comprehensive set of model variations, incorporating variations in resolution,  cosmological parameters, and baryonic feedback implementations. These model variations allow for comprehensive analysis of the sensitivity of WL observables to galaxy formation physics, especially in the context of high-precision surveys like Euclid and LSST.

The main parameters of all simulations analysed in this work are summarised in Table~\ref{tab:sim_variations}. The simulations are labelled according to their box size in comoving Gpc (cGpc) and base-10 logarithm of mass resolution. To illustrate, `L1\_m8'  refers to a run with a box size of 1 cGpc on the side and the baryonic particle mass of $\sim 10^8\:$\(\textup{M}_\odot\).

The set includes four simulations with different resolutions and box sizes calibrated to the same astrophysical parameters and the fiducial $\Lambda$CDM cosmological model from the DES Year 3 \citep[$3\times2$pt + All Ext.; ][]{PhysRevD.105.023520}, which is referred to as fiducial cosmology and denoted by `D3A'. The run of a $(1\,\mathrm{cGpc})^3$ volume has baryonic particle mass of $\sim 10^9\:$\(\textup{M}_\odot\). The flagship run  L2p8\_m9 with a volume of $(2.8\,\mathrm{cGpc})^3$ and baryonic particle mass of $\sim 10^9\:\mathrm{M}_\odot$, has in total approximately $3 \times 10^{11}$ particles.

Nine simulations performed in the $(1\,\mathrm{cGpc})^3$ volume assume the fiducial cosmology but vary stellar and/or AGN feedback. These feedback models were calibrated to match different combinations of galaxy SMF ($M^*$) and cluster gas fractions ($f_{\rm gas}$), using either thermal or kinetic AGN feedback modes. The fiducial feedback model reproduces the $z=0$ SMF measured by \cite{Driver_2022} for the GAMA survey and cluster gas fractions from low-redshift X-ray observations compiled in \cite{2023MNRAS.526.6103K}. To explore deviations from this benchmark, FLAMINGO has four runs, labelled as $f_{\text{gas}} \pm n\sigma$, which change only gas fractions by integer multiples of their observational scatter $\sigma$ extracted by \cite{2023MNRAS.526.6103K}. The $M^* -\sigma$ model lowers the SMF by 0.14 dex, which is consistent with the expected systematic uncertainty from \cite{Behroozi_2019}.
The $M^*-\sigma\_\text{fgas}-4\sigma$ combines both SMF and gas fraction shifts. All of these models apply thermal AGN feedback, while two variations marked with `Jet' apply kinetic jet feedback, facilitating a direct comparison of feedback implementations calibrated to the same observational targets. The box sizes, resolutions, and other simulation characteristics are listed in  Table~\ref{tab:sim_variations}.

\begin{table*}
\centering
\caption{Summary of FLAMINGO simulation parameters for runs with the box sizes of $\rm L=$ 1 cGpc and $N_{\rm p}=1800^3$ baryonic and dark matter particles. Columns indicate: simulation identifier; baryonic mass resolution, $m_{\rm b}$ ($10^9\, \rm M_\odot$); calibration shifts in galaxy SMF, $\Delta \rm M_*$, and cluster gas fraction, $\Delta f_{\rm gas}$, in units of observational standard deviation $\sigma$; AGN feedback mode; dimensionless Hubble parameter, $h$; present-day matter and baryon density fractions, $\Omega_{\rm m}$ and $\Omega_{\rm b}$; sum of neutrino masses, $\sum m_\nu c^2$ (eV); present-day linear rms fluctuation in 8 Mpc $h^{-1}$ spheres, $\sigma_8$; and the parameter defining the amplitude of the initial power spectrum  $S_8 = \sigma_8 \sqrt{\Omega_\mathrm{m}/0.3}$.}

\label{tab:sim_variations}
\resizebox{\textwidth}{!}{%
\begin{tabular}{lcccccccccc}
\hline
Identifier  & $m_{\rm b}$  & $\Delta \rm M_*$  & $\Delta f_{\mathrm{gas}}$  & AGN & $h$ & $\Omega_{\rm m}$ & $\Omega_{\rm b}$ & $\sum m_\nu c^2$  & $\sigma_8$ & $S_8$ \\
 & ($ 10^9\, \rm M_\odot$) & ($\sigma$) & ($\sigma$) &   &   &   &   & (eV) &   &   \\

\hline

L1\_m9                     & 1.07 & 0 & 0 & thermal & 0.681 & 0.306 & 0.0486 & 0.06 & 0.807 & 	0.815 \\
Planck                    & 1.07 & 0 & 0 & thermal & 0.673 & 0.316 & 0.0494 & 0.06 & 0.812 & 0.833 \\
PlanckNu0p24Var           & 1.06 & 0 & 0 & thermal & 0.662 & 0.328 & 0.0510 & 0.24 & 0.772 & 0.807 \\
PlanckNu0p24Fix           & 1.07 & 0 & 0 & thermal & 0.673 & 0.316 & 0.0494 & 0.24 & 0.769 & 0.789 \\

LS8                       & 1.07 & 0 & 0 & thermal & 0.682 & 0.305 & 0.0473 & 0.06 & 0.760 & 0.766 \\
\hline
fgas+2$\sigma$             & 1.07 & 0 & +2 & thermal & 0.681 & 0.306 & 0.0486 & 0.06 & 0.807 & 0.815 \\
fgas--2$\sigma$            & 1.07 & 0 & --2 & thermal & 0.681 & 0.306 & 0.0486 & 0.06 & 0.807 & 0.815 \\
fgas--4$\sigma$            & 1.07 & 0 & --4 & thermal & 0.681 & 0.306 & 0.0486 & 0.06 & 0.807 & 0.815 \\
fgas--8$\sigma$            & 1.07 & 0 & --8 & thermal & 0.681 & 0.306 & 0.0486 & 0.06 & 0.807 & 0.815 \\
M$*-\sigma$                & 1.07 & --1 & 0 & thermal & 0.681 & 0.306 & 0.0486 & 0.06 & 0.807 & 0.815 \\
M$*-\sigma$\_fgas--4$\sigma$   & 1.07 & --1 & --4 & thermal & 0.681 & 0.306 & 0.0486 & 0.06 & 0.807 & 0.815 \\
Jet                        & 1.07 & 0 & 0 & jet     & 0.681 & 0.306 & 0.0486 & 0.06 & 0.807 & 0.815 \\
Jet\_fgas--4$\sigma$       & 1.07 & 0 & --4 & jet     & 0.681 & 0.306 & 0.0486 & 0.06 & 0.807 & 0.815 \\

\hline
\end{tabular}%
}
\end{table*}


In addition to feedback variations, the FLAMINGO suite includes a set of five cosmology variations run in the same $(1\,\mathrm{cGpc})^3$ box with the fiducial baryonic model. Among them is the fiducial D3A model. Three other cosmologies are derived from the Planck best-fit $\Lambda$CDM model \citep{2020A&A...641A...6P}: one with a single massive neutrino of $m_\nu = 0.06$ eV, another two assume a higher total neutrino mass of $\sum m_\nu = 0.24$ eV (three species with $m_\nu = 0.08$ eV) with alternative strategies for handling remaining cosmological parameters. One applies the parameters inferred from Planck posteriors, while the other keeps all of them fixed except $\Omega_{\rm CDM}$. Another cosmology variation uses a lower $S_8$ value, extracted from WL measurements combined with CMB by \citet{Amon_2022}. The relevant cosmological parameters for each variation are provided in Table~\ref{tab:sim_variations}.

To eliminate the biases from cosmic variance, all $(1\,\mathrm{cGpc})^3$ simulations were initialised with the same random seed and observer's positions. Moreover, each cosmology variation has a DMO counterpart, which enables the construction of transfer functions for baryonic suppression of the ST coefficients.


\subsection{Light-cones and WL Convergence Maps}
\label{subsec:maps}

A key feature of FLAMINGO simulations is the on-the-fly construction of particle light-cones. For each virtual observer, the past light-cone is subdivided into concentric spherical shells, where particles crossing the light-cone are projected and stored as HEALPix maps \citep{2005ApJ...622..759G} for various physical quantities (see Appendix A2 of \citealt{2023MNRAS.526.4978S} for details). The $(1\,\mathrm{cGpc})^3$ simulations include a single observer with 60 shells of uniform redshift spacing $\Delta z=0.05$, spanning from $z=0$ to $z=3$.


The construction of WL convergence maps is described in detail in~\citet{2024MNRAS.529.2309B}. Here, we briefly outline the key steps. First, for compatibility with HEALPY \citep{2019JOSS....4.1298Z}, the FLAMINGO mass maps were downsampled to $N_\text{side} = 8192$, corresponding to an angular resolution of 0.43 arcmin. Since the simulation box size of $1\,\mathrm{cGpc}$ does not extend to the full redshift range required for Stage IV WL surveys ($z = 3$), the boxes were replicated along the line of sight. To avoid artificial structure repetition, mass shells exceeding the box size were randomly rotated.

Each shell $n$ is converted to an overdensity map $\delta^n(\boldsymbol{\theta})$:
\begin{equation}
\delta^n(\boldsymbol{\theta}) = \frac{\Sigma^n(\boldsymbol{\theta}) - \overline{\Sigma^n}}{\overline{\Sigma^n}},
\label{eq:delta}
\end{equation}
where $\Sigma^n(\boldsymbol{\theta})$ is the projected surface mass density at angular position $\boldsymbol{\theta}$ and $\overline{\Sigma^n}$ is its mean. This is used to compute the convergence $K^{n}$ at each lens plane $n$:
\begin{equation}
    K^n(\boldsymbol{\theta})  = \frac{3H_0^2 \Omega_\text{m}}{2c^2}\chi^n (1+z^n) \Delta\chi^n\delta^{n}(\boldsymbol{\theta}) \equiv \nabla^2 \psi^{n},
    \label{eq:poisson}
\end{equation}
where $\chi^n$, $z^n$, $\Delta\chi^n$, and $\psi^n$ are the comoving distance, redshift, comoving thickness, and lensing potential of the shell, respectively. The latter is obtained by solving the Poisson equation in harmonic space:
\begin{equation}
    \psi^n_{\ell m} = -\frac{2 K^n_{\ell m}}{\ell (\ell +1)},
\end{equation}
and then used to derive deflection angles and shear matrices. The resulting magnification matrix is propagated across shells, updating photon trajectories and recalculating convergence at each step. 

The final convergence field $\kappa(\boldsymbol{\theta})$ is constructed by integrating lensing contributions weighted by a source redshift distribution:
\begin{equation}
    \kappa(\boldsymbol{\theta}) = \int_0^{\chi_\text{hor}} \mathrm{d}\chi n[z(\chi)]\kappa(\boldsymbol{\theta},\chi), 
    \label{eq:kappa}
\end{equation}
where $\chi_{\rm hor}$ corresponds to $\chi(z=3)$, and $n(z)$ is the normalized redshift distribution of source galaxies.

In this paper, we assumed a Euclid-like source redshift distribution \citep{2020A&A...642A.191E}:
\begin{equation}
n_s(z) \propto \left(\frac{z}{z_0}\right)^2 \exp\left[ - \left( \frac{z}{z_0} \right)^{3/2} \right], \qquad z_0 = 0.9/\sqrt{2}.
\label{eq:nz}
\end{equation}

Although this choice is survey-specific, we assume that small variations in $n(z)$ have a negligible impact on the resulting HOS.  Nonetheless, significant changes in $n(z)$ can lead to noticeably different convergence maps. 

\section{Scattering transform}

\label{sec:scattering_transform}

In this paper, we apply the ST, an approach that combines wavelet analysis and CNNs. Initially introduced for signal processing and image classification \citep{2012arXiv1203.1513B, 2011arXiv1101.2286M}, the ST has found successful applications in science, including astrophysics \citep[e.g.,][]{2019A&A...629A.115A, saydjari_classification_2021, 2021A&A...649L..18R, 2022MNRAS.517.1625C} and cosmology, namely  WL \citep[e.g.,][]{2020MNRAS.499.5902C, 2021arXiv211201288C, 2021MNRAS.507.1012C}, LLS~\citep[e.g.,][]{2022PhRvD.105j3534V, 2022PhRvD.106j3509V, 2024PhRvD.109j3503V,2024JCAP...11..061V,  2024JCAP...07..021P, 2024PhRvD.109h3535R}, and IGM~\citep[e.g.,][]{2022MNRAS.513.1719G, 2024PhRvL.132w1002T}. It has shown its potential to extract non-Gaussian features from LSS data \citep[e.g.,][]{2025JCAP...01..006C, 2024PhRvD.110j3539G}.

Although the scattering transform has no relation to the scattering processes in physics, one can use the following analogy for an intuitive understanding of the method. An input signal is "scattered" through cascades of wavelet convolutions, extracting information across different scales and orientations at each step. This hierarchical process resembles a CNN with pre-defined filters. 
Just like CNNs, the ST captures spatial features through sequential layers of local convolutions and non-linear operations. However, since it does not require training, the ST is significantly more computationally efficient and interpretable in terms of scale, shape, and orientation.

From a statistical perspective, the ST has similarities with $N$-point correlation functions \citep[e.g.,][]{2002A&A...389L..28B}, because it captures complex patterns beyond just averages and variances. At the same time, it avoids noise amplification and the dilution of information across a large number of descriptors by applying a modulus as a non-linear transformation instead of field intensity multiplication. The result is a compact set of scattering coefficients that efficiently capture the non-Gaussian features (e.g., sparsity and clustering) of the input field, while being insensitive to outliers, noise and cosmic variance. These properties make the ST a perfect tool for analysing  WL convergence maps \citep{2021MNRAS.507.1012C}. 

In this paper, we use the publicly available ST routine developed by \cite{2020MNRAS.499.5902C}.\footnote{\href{https://github.com/SihaoCheng/scattering_transform/tree/master/scattering}{https://github.com/SihaoCheng/scattering\_transform}} Below, we summarise the key ideas and formal definitions relevant to our analysis, for a detailed overview see \cite{2021arXiv211201288C}. 

The ST is based on two repeatedly applied operations: a wavelet convolution ($\star$) and a pixel-wise modulus ($|\cdot|$). For an input field $I_0(x,y)\equiv I_0(\boldsymbol{x})$ and a family of complex-valued oriented wavelets $\psi^{j,l}(\boldsymbol{x})$ indexed by scale $j$ and orientation $l$, the first-order field $I_1^{j_1,l_1}(\boldsymbol{x})$ is:
\begin{equation}
I_1^{j_1,l_1}(\boldsymbol{x}) = \left| I_0(\boldsymbol{x}) \star \psi^{j_1,l_1}(\boldsymbol{x}) \right|.
\end{equation}

The intuition behind these operations is the following: convolution highlights structures at a certain scale and orientation, and the modulus gives the amplitude of the fluctuation.

The second-order fields are generated in the same way:

\begin{equation}
I_n^{j_1,l_1,j_2,l_2}(\boldsymbol{x}) = \left| \left| I_0(\boldsymbol{x}) \star \psi^{j_1,l_1}(\boldsymbol{x}) \right| \star \psi^{j_2,l_2}(\boldsymbol{x}) \right|.
\end{equation}

Higher-order terms can be computed similarly by continuing this cascade. However, higher-order coefficients rapidly decay in magnitude and provide little additional information \citep[e.g.,][]{2020MNRAS.499.5902C, 2016arXiv160507464W}. In this paper, we compute the ST only up to the second order. 

The scattering coefficients $\mathcal{S}_n$ are then obtained by computing the expected value of the transformed fields, which for homogeneous fields is the same as the spatial average ($\langle\cdot\rangle_{\boldsymbol{x}}$)

\begin{align}
\mathcal{S}_0 &= \langle I_0(\boldsymbol{x}) \rangle_{\boldsymbol{x}}, \\
\mathcal{S}_1^{j_1,l_1} &= \left\langle \left| I_0 \star \psi^{j_1,l_1} \right| \right\rangle_{\boldsymbol{x}}, \\
\mathcal{S}_2^{j_1,l_1,j_2,l_2} &= \left\langle \left| \left| I_0 \star \psi^{j_1,l_1} \right| \star \psi^{j_2,l_2} \right| \right\rangle_{\boldsymbol{x}}.
\end{align}

These translation-invariant descriptors have a clear physical interpretation. $\mathcal{S}_0$ is the average value of the field, $\mathcal{S}_1^{j_1,l_1}$ are analogous to the power spectrum at the wavelet central frequency, and $\mathcal{S}_2^{j_1,l_1,j_2,l_2}$ measures clustering of $ j_1$-scale structures on $j_2$ scales, capturing correlations between scales. 

The wavelets $\psi^{j,l}(\boldsymbol{x})$ are produced by rotating and dilating a mother wavelet. The dilation factor $2^j$ determines the number of pixels occupied by a wavelet
 \begin{equation}
 \psi^{j,l}(\boldsymbol{x}) = 2^{-2j} \cdot \psi\left( 2^{-j}r_l^{-1} \boldsymbol{x}\right),
 \end{equation}
 where $r_l$ denotes a rotation by angle $\pi\: l/L$ with $L$ being the total number of discrete angular orientations and $l$ varying from $1$ to $L$.
 
In the Fourier domain, this corresponds to
 \begin{equation}
 \hat{\psi}^{j,l}(\boldsymbol{k}) = \hat{\psi}\left( 2^{j}r_l^{-1} \boldsymbol{k}\right).
 \end{equation}

For the completeness and stability of the ST, the wavelet filters must be localised in both real and frequency domains, cover the Fourier space uniformly (except zero frequency), and be stable to small deformations in the input.

The standard choice that satisfies these conditions is the Morlet wavelet~\citep[][]{2020MNRAS.499.5902C}. Although this paper presents only the results for Morlet wavelets, we also generated transfer functions for other wavelet families, such as Gaussian, Gaussian harmonic, bump steerable, and Shannon, all of which are represented in Appendix~\ref{appendix:wavelets}. 

For $J$ dyadic scales with $L$ orientations, there are $1 + J L + J^2 L^2$ scattering coefficients. However, as shown in \cite{2021MNRAS.507.1012C}, for isotropic fields such as  WL convergence maps, orientation dependence can be reduced by averaging over $l_1$ and $l_2$. Moreover, of the 2nd-order coefficients, only those with $j_1<j_2$ are informative.
This results in a significantly reduced set

\begin{align}
\mathcal{S}_1^{j_1} &= \langle \mathcal{S}_1^{j_1,l_1} \rangle_{l_1}, \\
\mathcal{S}_2^{j_1,j_2} &= \langle \mathcal{S}_2^{j_1,l_1,j_2,l_2} \rangle_{l_1,l_2},
\end{align}
with $j_1 < j_2$. 

Then, the total number of coefficients is $1 + J + J(J+1)/2$. For $512 \times 512$ pixel maps and $J=8$ scales, this gives $1 + 8 + 28 = 45$ coefficients.

\section{Results}

\label{sec:results}

\begin{figure}
    \centering

    \begin{minipage}{0.48\linewidth}
        \centering
        \includegraphics[width=\linewidth]{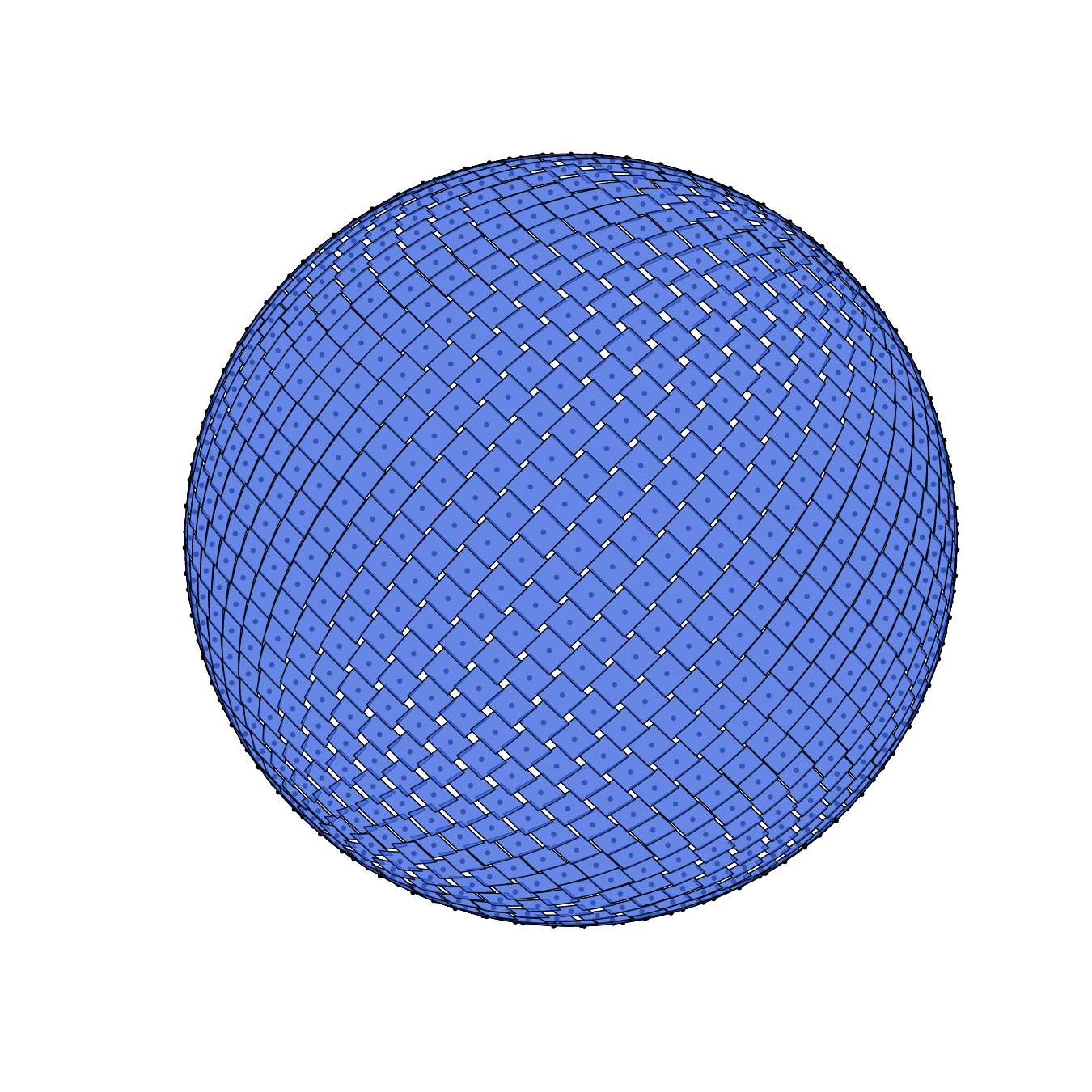}
    \end{minipage}
    \hfill
    \begin{minipage}{0.48\linewidth}
        \centering
        \includegraphics[width=\linewidth]{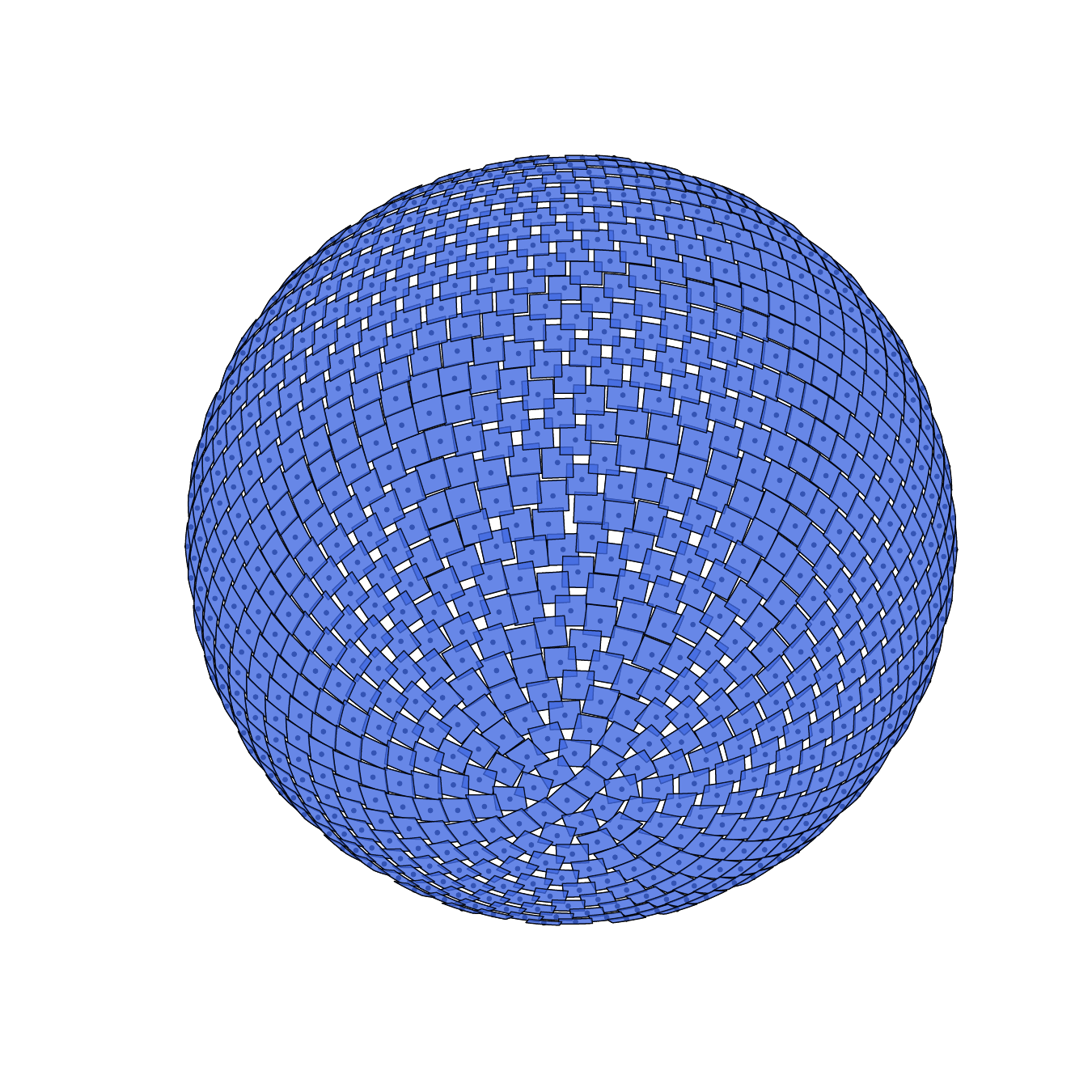}
    \end{minipage}
    \caption{Example of the full-sphere coverage by $5\times5$ $\mathrm{deg}^2$ patches, with their centres distributed using a Fibonacci lattice. The left figure shows the sphere viewed from the equatorial direction ($x>0$, corresponding to $\varphi = 0$), while the right figure shows the view from the pole ($z>0$, $\theta = 0$). The patches are nearly non-overlapping and sample the sphere approximately uniformy.}
    \label{fig:Fibonacci}
\end{figure}

\begin{figure*}
    \centering
    \includegraphics[width=0.9\textwidth]{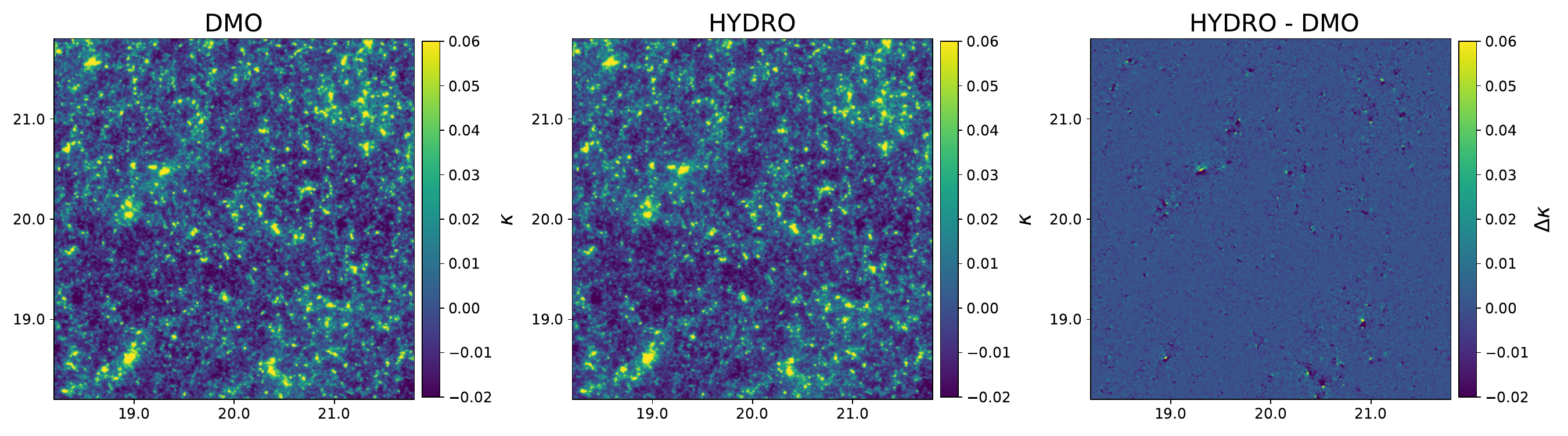}
    \caption{Convergence map patches for the DMO (left) and hydrodynamical (center) variations of L1\_m9. Each patch covers $3.6 \times 3.6$ $\text{deg}^2$ and includes structures up to $z = 3$ with a Euclid-like source redshift distribution. The right panel shows their difference, highlighting the non-Gaussian features created by baryonic effects. No noise or smoothing is applied. Maps are generated via backward ray-tracing, as described in Section~\ref{subsec:maps}. 
    }
    \label{fig:diff_structures}
\end{figure*}

\begin{figure*}
    \centering
    \includegraphics[width=0.9\textwidth]{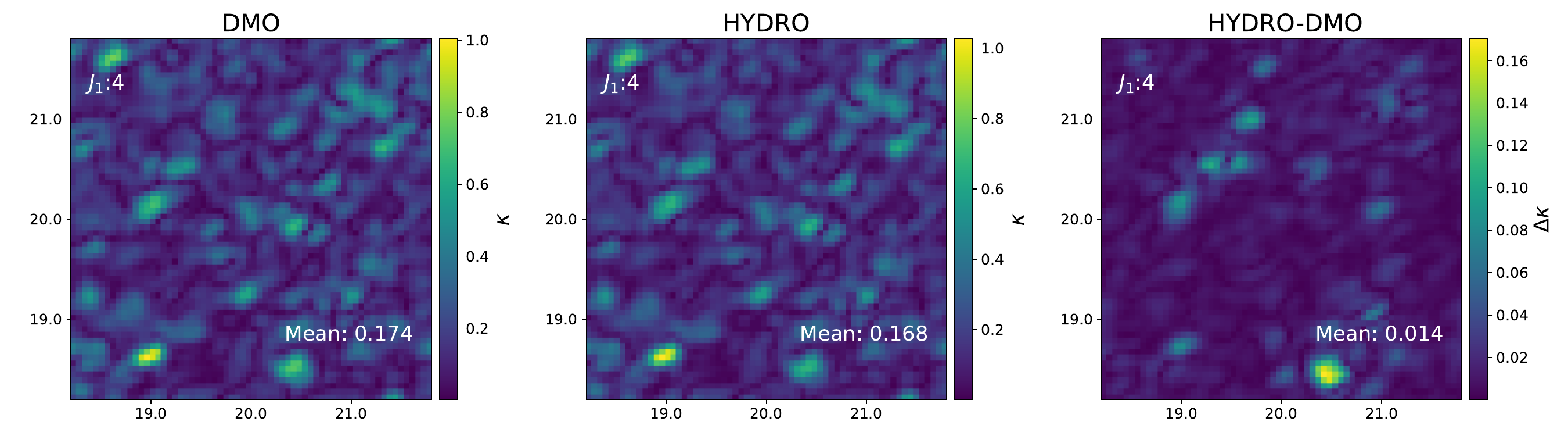}
    \caption{ Same as Figure~\ref{fig:diff_structures}, but showing the moduli of the convolutions with Morlet wavelets at $(j = 4, l = 2)$, with azimuthal resolution $L = 8$. The mean value of the modulus of each convolved field is shown in the bottom-right corner of each panel. The wavelets effectively capture structural differences caused by baryonic effects, as is evident from the comparison with the right panel of Figure~\ref{fig:diff_structures}: the regions of largest absolute values of $\Delta\kappa$ coincide, while areas with no difference remain near zero. Note that the third panel uses a different colour scale, which makes the smaller differences visible.}
    \label{fig:diff_structures_convolved}
\end{figure*}

To compute scattering coefficients, we divided the full-sky convergence maps into 3183 almost non-overlapping patches of $3.6 \times 3.6$ $\text{deg}^2$ with centes quasi-uniformly distributed on the sphere \footnote{For the analysis presented in this paper, we chose the patch size following~\cite{2020MNRAS.499.5902C}. We also computed the transfer functions for other patch sizes: $1.8 \times 1.8 \,\mathrm{deg}^2$, $5 \times 5 \,\mathrm{deg}^2$, and $7.5 \times 7.5 \,\mathrm{deg}^2$, which are available at \href{https://github.com/Mariia-Marinichenko/BF-ST-transfer-functions}{https://github.com/Mariia-Marinichenko/BF-ST-transfer-functions}.} using a Fibonacci lattice, a simple and efficient method that  is gaining popularity in  WL applications~\citep[e.g.,][]{2023MNRAS.524.5591F}. The Fibonacci lattice consists of a given number of evenly distributed  points on a unit sphere with coordinates 
\begin{equation}
    \{\theta, \varphi\} = \bigg\{\arccos\bigg(1 - \frac{2(n+0.5)}{N}\bigg) , \frac{2\, \pi \, n}{\phi}\bigg\}, \qquad 0\leq n \leq N.
\end{equation}
Here $\phi =(\sqrt{5}+1)/2\approx 1.618$ is golden ratio, and $N$ is the total number of points. Figure~\ref{fig:Fibonacci} illustrates the example coverage of the sphere. Each patch has a resolution of $512 \times 512$ pixels.
This procedure was done for both DMO and hydrodynamical simulations. As shown in Figure~\ref{fig:diff_structures}, thanks to the matched initial seeds and observer's positions, we can isolate the baryonic impact from cosmic variance, 
showing the scale-dependent baryonic influence on $\kappa$.  For each patch, we calculated the reduced scattering coefficients up to second order, using a maximum scale of $J=8$. Figure~\ref{fig:diff_structures_convolved} illustrates the same patches convolved with a Morlet wavelet with $(j = 4, l = 2)$, highlighting differences in structures. 

In the remainder of this paper, we refer to characteristic scale ranges using the definitions provided in Table~\ref{tab:scales}.

\begin{table}
    \centering
    \caption{Scale definitions used throughout the paper, with corresponding diadic scales $j$, Fourier wavenumbers $k$, and real-space scales $D$ at $z\approx1$.}
        \begin{tabular}{lccc}
        \hline
        Name & $j$ & $k$  & $D$  \\
         &  &($h\,\mathrm{Mpc}^{-1}$) &  ($\mathrm{Mpc}\,h^{-1}$) \\
        \hline
        Smallest      & $0$     & $3.6$             & $0.3$           \\
        Small         & $1$ – $2$ &  $1.8$ - $0.9$      & $0.6$ – $1 $     \\
        Intermediate  & $3$ – $5$ & $0.4$ – $0.1$      & $2$ – $9$       \\
        Large         & $6$ – $7$ & $0.06$ – $0.03$      & $18$ – $36$       \\
        \hline
    \end{tabular}
    \label{tab:scales}
\end{table}

\subsection{Scattering coefficients}
\label{subsec:results_ST}
\begin{figure*}
    \centering
    \includegraphics[width=\textwidth]{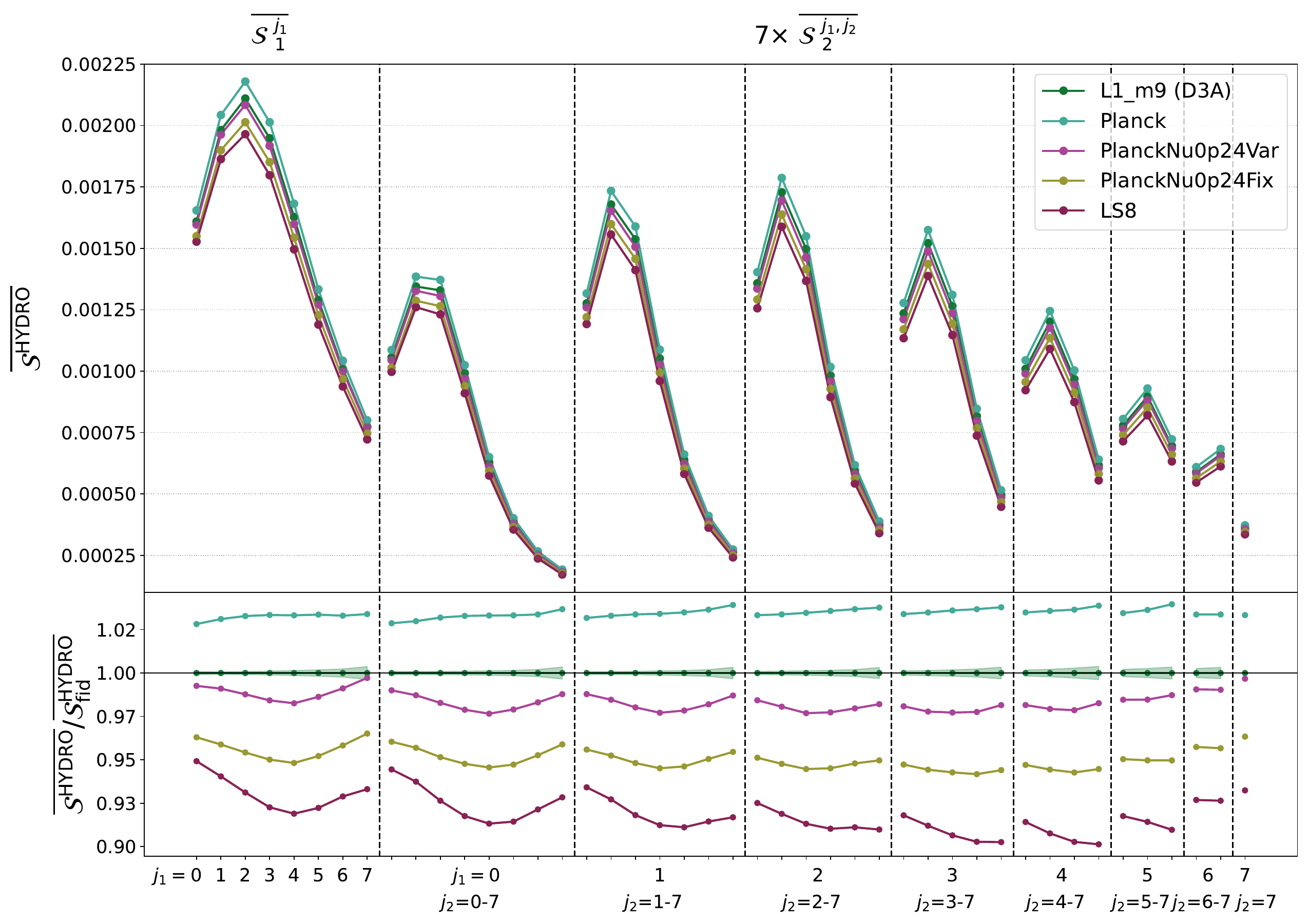}
    \caption{Scattering coefficients $\mathcal{S}$ as a function of dyadic scale across all FLAMINGO cosmologies, computed from noiseless, unsmoothed convergence maps and averaged over angular orientations and patches. The top panel shows absolute values of the first-order ($\mathcal{S}_1$) and second-order ($\mathcal{S}_2$) scattering coefficients. For illustrative purposes, $\mathcal{S}_2$ were multiplied by a factor of $7$. The $\mathcal{S}_1$ coefficients correspond to the amplitudes of $\kappa$ fluctuations and resemble the convergence power spectrum, while $\mathcal{S}_2$ describe cross-scale clustering. The lower panel displays ratios relative to the fiducial D3A model (L1\_m9).  The green-shaded region corresponds to the cosmic variance in L1\_m9. Cosmological variations affect all coefficients uniformly, leaving an approximately scale-independent imprint. }
    \label{fig:sc_all_cosmologies}
\end{figure*}

Figure~\ref{fig:sc_all_cosmologies} displays the reduced scattering coefficients for all FLAMINGO cosmologies listed in Table~\ref{tab:sim_variations}, computed from noiseless, unsmoothed maps.  The coefficients have been averaged over all angular orientations $ l \in [0, L]$ and patches (i.e. realisations). The first- and second-order scattering coefficients are presented as a function of increasing dyadic scales ($j_1 \in [0, J]$, $j_2 \in [j_1, J]$ with $J=7$).

The top panel shows first- and second- order scattering coefficients, grouped by scale. The first-order coefficients, which reflect the amplitude of $\kappa$ fluctuations, resemble the convergence power spectrum due to their similar physical meanings. The second-order coefficients, shown for each $j_1$, characterise cross-scale correlations~\citep{2020MNRAS.499.5902C} and have the same shape, but with a smaller amplitude due to the additional wavelet convolution.  The lower panel shows the ratio of these coefficients to those from the fiducial D3A cosmology, where the shaded regions correspond to the standard deviation of the fiducial coefficients. As expected, cosmic variance, estimated as the standard error of the mean over all patches and represented by a green-shaded region, increases on larger scales, but the models remain well separated across all scales.

In general, scattering coefficients demonstrate strong sensitivity to the parameter $S_8 = \sigma_8 \sqrt{\Omega_\mathrm{m}/0.3}$, consistent with their equivalence to the power spectrum, which is also sensitive to $S_8$~\citep{kilbinger_cosmology_2015}. 
As a result, they decrease with lower $S_8$. For example, the LS8 cosmology, which has an $8\%$ lower $S_8$ than the Planck model, leads to a global amplitude reduction exceeding $10\,\%$, consistent with other lensing observables such as the Sunyaev-Zel'dovich effect (SZE) $y$ – CMB lensing $\kappa$ cross-spectrum \citep[][their Figure 17]{2023MNRAS.526.4978S}.
A detailed analysis of the functional dependence of the scattering coefficients on $S_8$ is provided in Appendix~\ref{appendix:S_8_dependence}. 

\begin{figure*}
    \centering
    \includegraphics[width=\textwidth]{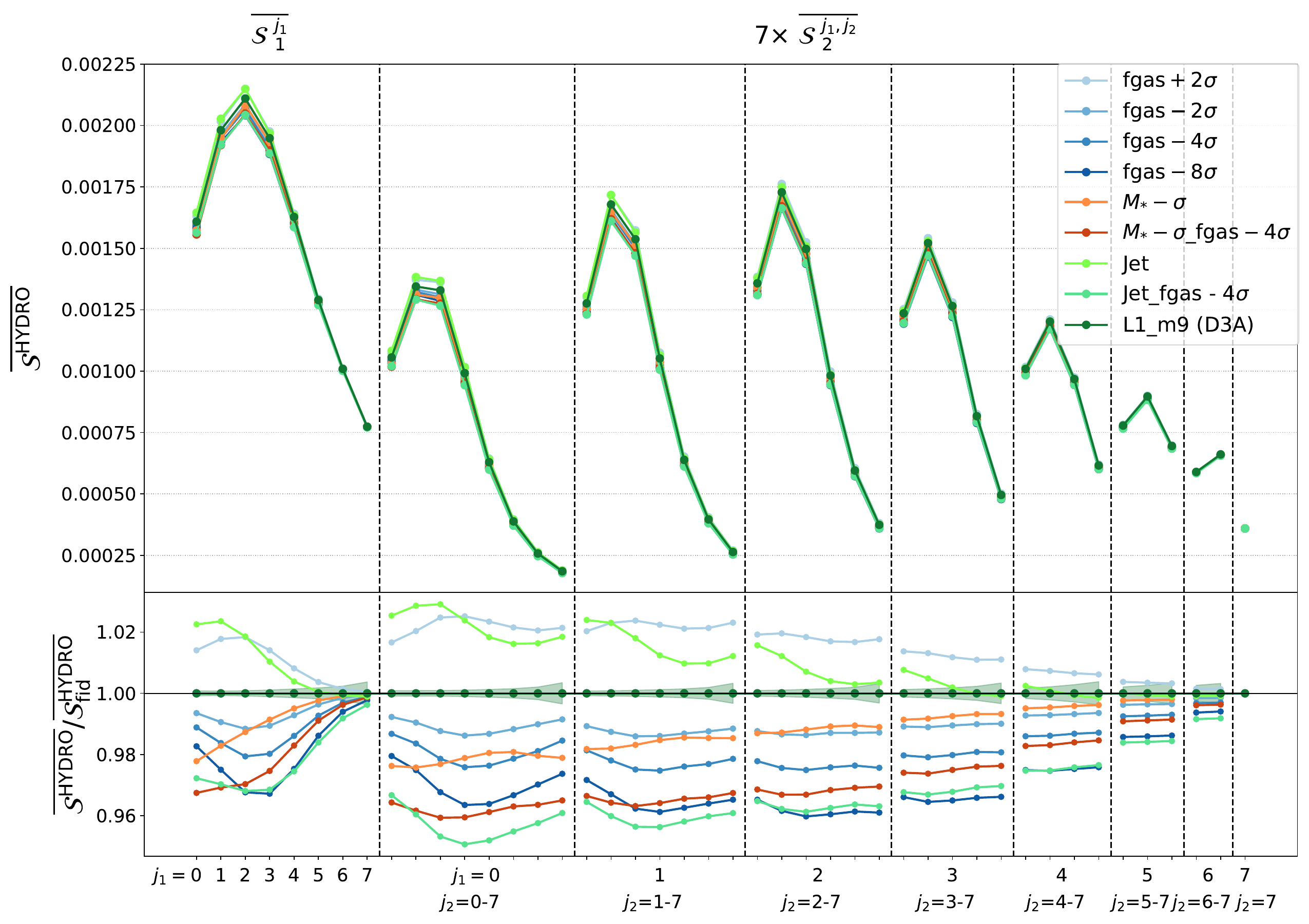}
    \caption{Scattering coefficients $\mathcal{S}$ as a function of dyadic scale for the D3A cosmology across all FLAMINGO baryonic feedback models, computed from noiseless, unsmoothed convergence maps and averaged over angular orientations and patches. The top panel shows absolute values of the first-order ($\mathcal{S}_1$) and second-order ($\mathcal{S}_2$) coefficients. For illustrative purposes, $\mathcal{S}_2$ were multiplied by a factor of $7$. The $\mathcal{S}_1$ coefficients correspond to the amplitudes of $\kappa$ fluctuations and resemble the convergence power spectrum, while $\mathcal{S}_2$ describe cross-scale clustering. The lower panel displays ratios relative to the fiducial feedback model (L1\_m9). The green-shaded region corresponds to the cosmic variance in the fiducial model. Baryonic feedback introduces scale-dependent suppression that strongly affects small scales.}
    \label{fig:sc_all_feedbacks}
\end{figure*}

Baryonic feedback, even within a fixed cosmology, introduces comparable discrepancy. Figure~\ref{fig:sc_all_feedbacks} shows the scattering coefficients for the D3A cosmology across all FLAMINGO baryonic feedback models. Between the two most extreme feedback scenarios, where cluster gas fractions differ by $10\sigma$, differences in scattering coefficients reach $\sim 8\,\%$. This implies that baryonic effects leave an imprint nearly as significant as cosmology does, and must be carefully accounted for in high-precision analyses.

Importantly,  while cosmology affects all coefficients uniformly (through $S_8$ rescaling), baryonic feedback is scale-dependent and primarily suppresses small structures. This split in behaviour allows for statistical separation of cosmological and baryonic contributions. Advanced techniques such as Bayesian model averaging \citep[e.g.,][]{2025MNRAS.536.2064G} can be applied to marginalise over feedback uncertainties and unbias cosmological inference.

To assist such methods, we next present the full transfer functions across all coefficients and models. 

\subsection{Transfer function}
\label{subsec:results_TF}

We demonstrate that the scattering transform retains a stable and predictive mapping between DMO and hydrodynamical simulations through a \textit{baryonic transfer function}, which we define as the ratio of average  scattering coefficients:
\begin{equation}
   \mathcal T =\frac{ \overline{\mathcal{S}^{\mathrm{HYDRO}}} }{ \overline{ \mathcal{S}^{\mathrm{DMO}}}},
\end{equation}
where overline denote the ensemble average over the patches. We also show that $\cal T$ is well-defined, nearly cosmology-independent, and vary for different baryonic feedback strength and mode. 

\begin{figure*}
    \centering
    \begin{subfigure}{\textwidth}
        \centering
        \includegraphics[width=0.85\linewidth]{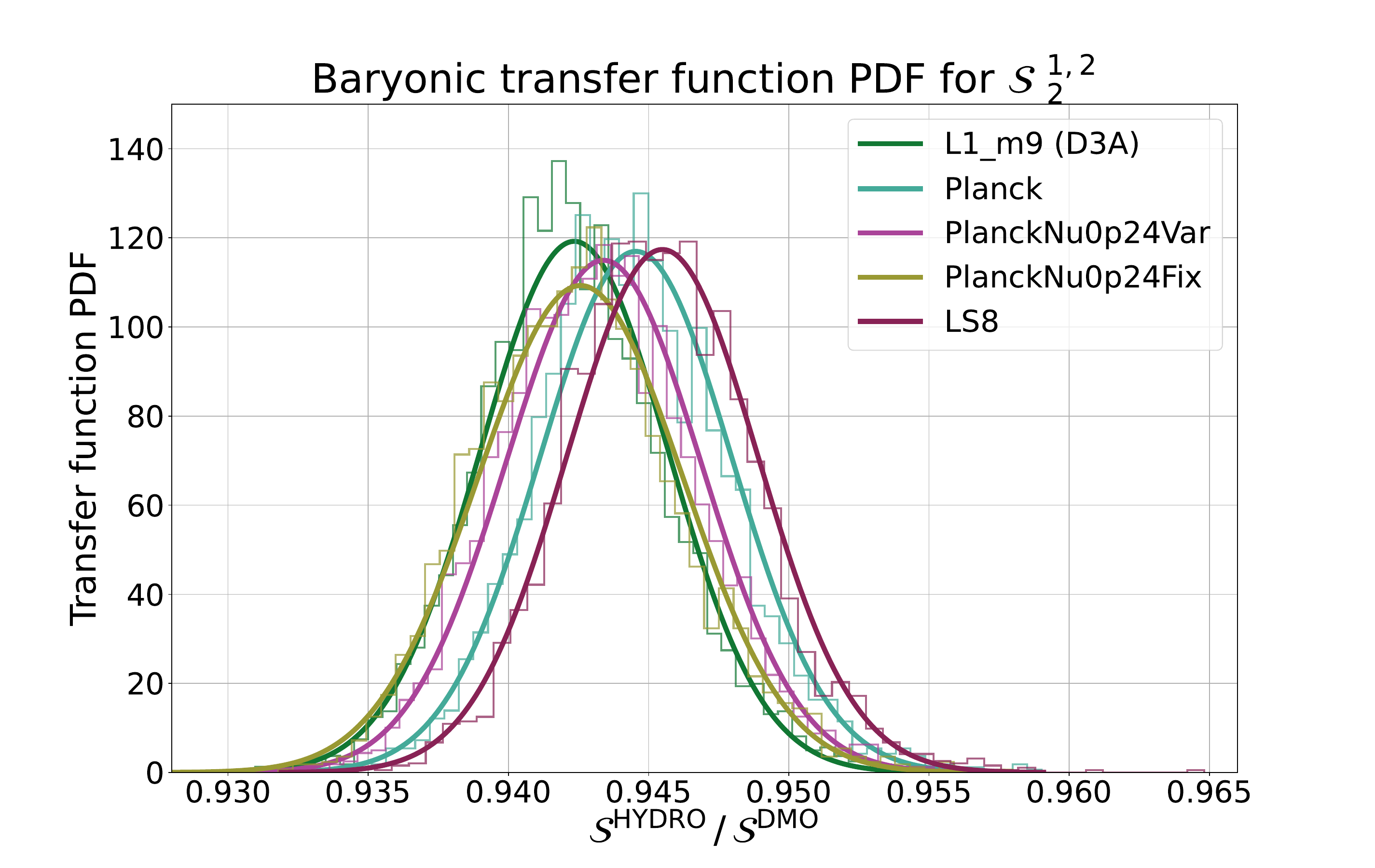}
        \caption{PDFs of transfer function for $j_1=1$, $ j_2=2$ (i.e., $\mathcal {T}^{1,2}_2$) for different cosmologies. Thick solid lines show Gaussian fits, while steps show the original distribution.}
        \label{fig:TF_hist_cosm}
    \end{subfigure}
    \vspace{-0.1mm}
    \begin{subfigure}{\textwidth}
        \centering
        \includegraphics[width=0.85\linewidth]{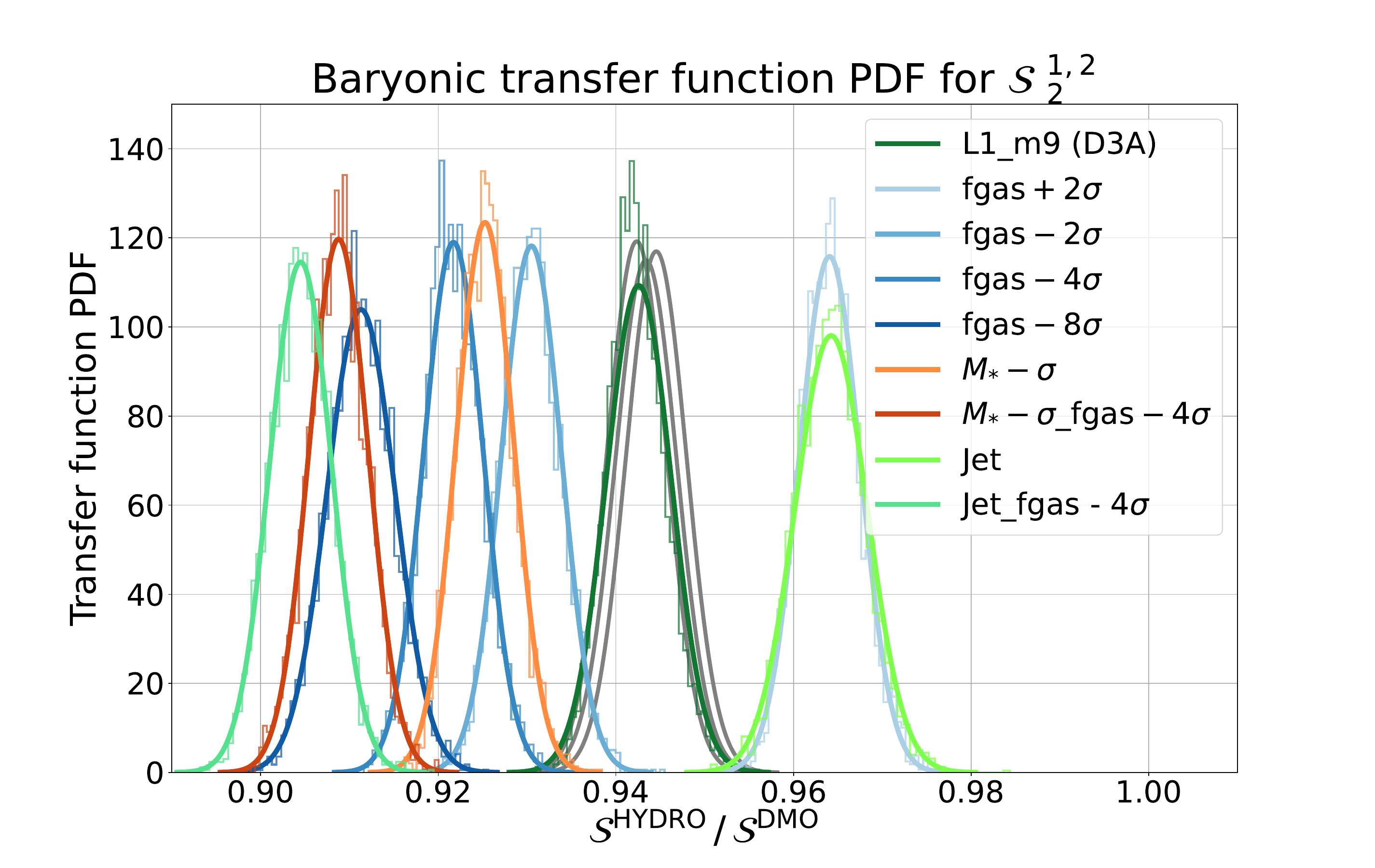}
        \caption{PDFs of transfer function for $j_1=1$, $ j_2=2$ (i.e., $\mathcal {T}^{1,2}_2$) for various feedback models. Grey curves reproduce (a) for comparison.}
        \label{fig:TF_hist_feedback}
    \end{subfigure}
    \caption{ PDFs of selected transfer function components $\mathcal {T}^{1,2}_2$ across patches, with stepped lines showing measured distributions and solid curves representing Gaussian fits. The top panel shows cosmology variations at fixed feedback, indicating minimal spread with means differ by $\lesssim 0.3\,\%$ and relative standard deviations are $\sim 0.3\,\%$. Panel (b) shows feedback variations at fixed cosmology (D3A), where means differ by up to $4\,\%$ and relative standard deviations reach $\sim 0.4\,\%$. Grey lines in (b) are the cosmological PDFs from (a), shown for reference. These results demonstrate that $\mathcal{T}$ is approximately Gaussian with small variance across patches, and is strongly sensitive to baryonic feedback while being roughly insensitive to cosmology. 
    }
    \label{fig:TF_hist_combined}
\end{figure*}

Before analysing the detailed structure of $\mathcal{T}$, we first examine the distribution of baryonic transfer ratios across all patches. Figure~\ref{fig:TF_hist_combined} shows the corresponding PDFs of the ratios for the $j_1 = 1$, $j_2 = 2$ scattering coefficients ($\mathcal {S}^{1,2}_2$), computed between patches at identical positions in the baryonic and DMO $\kappa$ maps (as in  Figure~\ref{fig:diff_structures}), together with Gaussian fits to these PDFs.

Cosmological variations (Figure~\ref{fig:TF_hist_cosm}) with the same fiducial feedback model show minimal spread: means differ by $\lesssim 0.3\,\%$, and the ratio of the standard deviation to mean is $\sim 0.3\,\%$. In contrast, the dispersion for different feedback implementations at fixed D3A cosmology (Figure~\ref{fig:TF_hist_feedback}) is significantly larger: means vary up to $4\,\%$, and the relative standard deviation $\sim 0.4\,\%$. Grey lines show the cosmological variation for reference.

These results establish two key findings:
\begin{itemize}
    \item the transfer function $\mathcal T$ is approximately normally distributed with small variance across patches;
    \item $\mathcal T$ is highly sensitive to feedback, while negligibly sensitive to cosmology.
\end{itemize}

This behaviour is consistent with the results of \citet{2024PhRvD.110j3539G} and supports the use of a simulation-based transfer function for correcting baryonic effects in cosmological inference.

Nevertheless, we note that we didn't explore large variations in cosmology.Substantial shifts in $S_8$, such as in LS8 run, may lead to noticeable differences in the transfer function.

To generalise our findings, we computed the full set of transfer functions across all scattering indices for various cosmologies and feedback models. 

\begin{figure*}
    \centering
    \includegraphics[width=\textwidth]{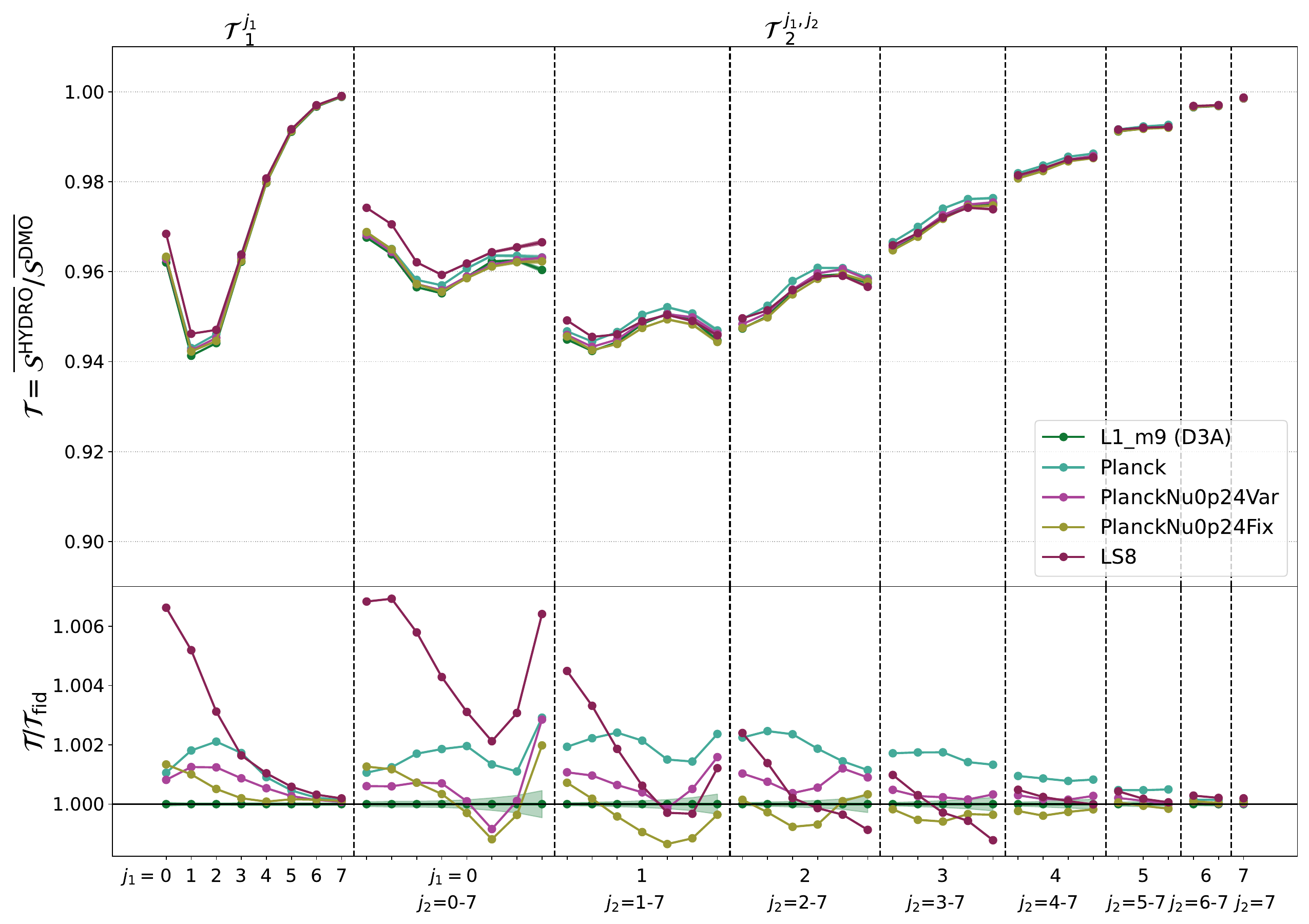}
    \caption{Transfer functions $\mathcal T^{j_1, j_2}_2$ as a function of dyadic scale across all FLAMINGO cosmologies, computed from noiseless, unsmoothed convergence maps and averaged over angular orientations and patches. The top panel shows absolute values of $\mathcal T^{j_1, j_2}_2$. The lower panel displays ratios relative to the fiducial D3A model (L1\_m9). The green-shaded region corresponds to the cosmic variance in the fiducial L1\_m9. Transfer functions demonstrate a characteristic "spoon-like" shape, with suppression strongest on small to intermediate scales ($j_1 = 1 -3$) and diminishing at larger scales ($j_1 \geq 5$). The curves are almost identical across cosmologies, confirming that $\mathcal{T}$ is insensitive to cosmology. The LS8 model shows the largest deviation, with reduced suppression on the smallest scale ($j = 0$), consistent with its lower $\sigma_8$ and a high-mass cutoff in the halo mass function shifted to lower halo masses relative to other models.  The shape of $\mathcal{T}^{j_1,j_2}_2$ for the second-order coefficients is largely inherited from the first-order $\mathcal{T}^{j_1}_1$, with clustering suppression saturating on $j_1 \geq 4$.
     }
    \label{fig:full_tf_cosmol}
\end{figure*}

\begin{figure*}
    \centering
    \includegraphics[width=\textwidth]{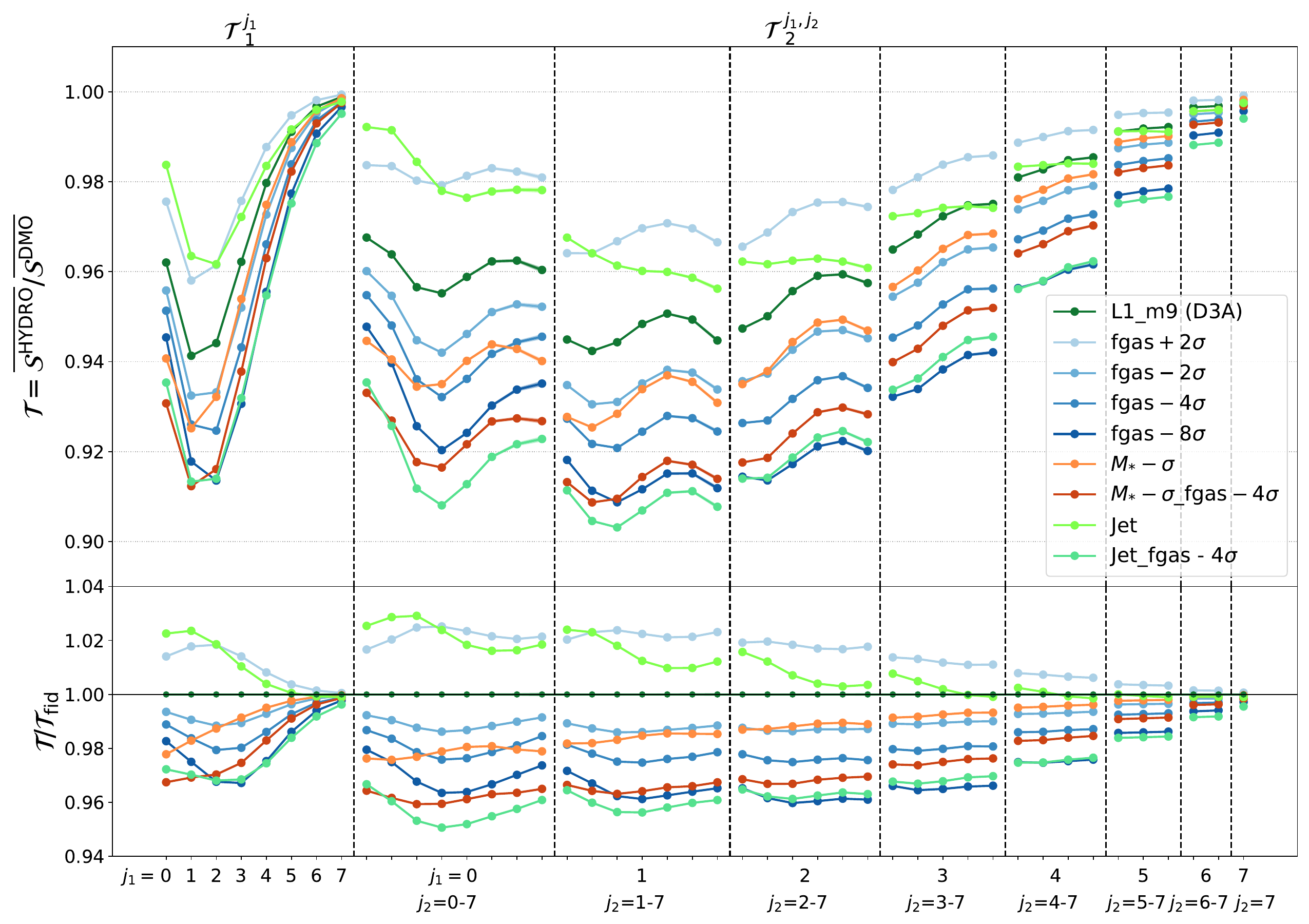}
    \caption{ Same as Figure~\ref{fig:full_tf_cosmol}, but for different baryonic feedback models. Baryonic effects introduce more significant and variable suppression compared to Figure~\ref{fig:full_tf_cosmol} (which has a much smaller range in the lower panel), especially on small to intermediate scales ($j_1 = 0-3$). On the smallest scale, $\mathcal{T}$ is enhanced relatively to larger $j$ due to gas cooling and star formation, while on larger scales, feedback is more severe, suppressing first-order coefficients by up to $9\%$. Variations in fgas produce non-linear responses, with stronger suppression requiring extreme $\sigma$ shifts. SMF variations produce greater small-scale suppression than fgas alone. Jet-mode AGN feedback leads to weaker small-scale suppression but stronger large-scale damping, reflecting differences in energy injection and spatial gas redistribution. When feedback becomes aggressive, as in Jet\_fgas - 4$\sigma$, suppression exceeds all other models on small scales and becomes indistinguishable from extreme kinetic feedback fgas - 8$\sigma$, highlighting the need for high-resolution data to distinguish feedback scenarios. 
    }
    \label{fig:full_tf_feedback}
\end{figure*}


\subsubsection{Different cosmologies}
Figure~\ref{fig:full_tf_cosmol} shows $\mathcal{T}$ across all coefficients and cosmologies. The curves are nearly identical, with the most noticeable deviation found in the  LS8 model. Each model produces a transfer function with a characteristic "spoon-like" shape similar to the power spectrum case. The suppression is strongest on small to intermediate scales ($j=1-3$), diminishing on larger scales, i.e. larger values of $j$. 

The transfer function for the second-order coefficients is largely determined by the damping of the first-filtered structures ($j_1$), i.e., the corresponding first-order transfer function $\mathcal{T}^{j_1}_1$.

On the smallest scales ($j = 0$), the transfer function is enhanced relative to larger scales due to baryonic clustering.
These scales ($~\sim 0.4$ arcmin or $\sim 0.3$  cMpc at $z = 1$) correspond to the regions where baryonic processes such as gas cooling and star formation concentrate matter in compact regions. On small scales ($j = 1-2$), feedback becomes most severe, reducing first-order scattering coefficients by approximately $6\,\%$. 
As scale increases, the impact of feedback diminishes and converges to $\sim1\,\%$ on large scales. In addition, for $j_1\geq 4$ the clustering suppression saturates and becomes independent of $j_2$.

Among cosmological variations, LS8 (with a $6\,\%$ lower $\sigma_8$) shows the least suppression on small scales ($j\leq 2$), differing from the fiducial one by $\sim1\,\%$. These discrepancies disappear with increasing scale, and all transfer functions agree with the fiducial model to within $1\%$ for $j_1\geq 5$. 
In addition, a relatively lower suppression of clustering on $j_1 = 0$ indicates that LS8 contains more small haloes. 
This is in line with expectations from the HMF. The lower $\sigma_8$ of LS8 with the $\Omega_\mathrm{m}$ identical to D3A shifts the high-mass cutoff of the halo mass function (HMF) to smaller halo masses \citep[e.g.,][]{Tinker_2008, 2024MNRAS.530.4203X}, resulting in fewer massive haloes with an even distribution of low-mass ones. 

For second-order coefficients, feedback significantly affects correlations on scales of $j_1 = 0-2$ and $j_2 = 2-4$, appearing as small "spoon-like" features.
In addition, the relative difference decreases up to  $j_2 = 5$ before increasing again. This may be attributed to the greater variance on the large scales. 

\subsubsection{Thermal feedback}

Figure~\ref{fig:full_tf_feedback} is analogous to Figure~\ref{fig:full_tf_cosmol}, but shows transfer functions across all FLAMINGO baryonic feedback implementations. Seven thermal AGN models were calibrated against different assumptions about the SMF, $M^*$ and cluster gas fraction, $f_\mathrm{gas}$. While the shape of the transfer function remains constant, its amplitude is sensitive to the feedback strength, demonstrating much larger differences than when changing cosmologies.

As was discussed before, on the smallest scales, the transfer function is relatively amplified, reflecting enhanced clustering from baryonic processes such as gas cooling and star formation that concentrate matter in compact regions.
This behaviour reverses towards intermediate scales of $j_1 = 2-3$,  reaching a stronger suppression of $4-9\,\%$. 



Interestingly, the response to changes in $f_{\rm gas}$ is non-linear. Increasing $f_{\rm gas}$ by $+2\sigma$ enhances the transfer function by approximately $2\,\%$ for individual structures ($\mathcal{S}^{j_1}_1$) and $3\,\%$ for clustering ($\mathcal{S}^{j_1,j_2}_2$)
, while a $-2\sigma$ reduction results in only a $\sim1\,\%$ decrease. To achieve a comparable $2 - 3\,\%$ suppression, the stronger shift of $-4\sigma$ is required, with a further reduction to fgas$-8\sigma$ producing only an additional $\sim1\,\%$ drop and indicating a saturation at extreme values. The suppression of density fluctuations peaks typically at $j = 2-3$, and is reflected in clustering. However, the latter is less sensitive, remaining relatively flat before gradually converging toward the fiducial model on larger scales. This non-linearity might arise from the scale-dependent nature of gas redistribution, which predominantly removes matter from structures on $\sim 1 \:\mathrm{Mpc}$ scales.

In contrast, varying the SMF at fixed gas fraction has a greater effect. A $-1\sigma$ shift in the SMF calibration (the $M_\star - \sigma$ model) leads to a $\sim\,2\,\%$ suppression in the transfer function on the small scales with the linear decline to the fiducial value. The impact on second-order coefficients is similar. When combined with a $-4\sigma$ shift in $f_{\rm gas}$, the suppression exceeds that of the fgas$-8\sigma$ model at $j = 0-1$, while $M_\star - \sigma$ alone already surpasses fgas$-8\sigma$ at $j = 0$. This suggests that changes in the SMF significantly affect the small scales, but have little effect on larger scales compared to the AGN feedback (i.e.\ lowering $f_{\rm gas}$). These results are in agreement with the HMF analysis of \cite{2023MNRAS.526.4978S} (their Figure 20), which shows the strongest suppression in SMF-shifted models at halo masses of $\sim 10^{11.5} \mathrm{M}_\odot  - 10^{13.5} \mathrm{M}_\odot $. This supports the idea that baryonic effects on the galaxy and cluster scale can be disentangled, a conclusion that also holds for the power spectrum, where these effects cause suppression on slightly different scales.

\subsubsection{Jet feedback}

Across scales corresponding to $j = 0-4$, the jet-mode AGN feedback produces a higher transfer function for first-order scattering coefficients than the fiducial kinetic model, indicating less efficient suppression of the small-scale structure, despite both being calibrated for the same SMF and cluster gas fractions. 
These differences indicate that jets may deposit their energy on larger scales, resulting in more extended but softer suppression. This is consistent with the higher gas fractions predicted for haloes in the $\sim10^{13}$–$10^{14} \, \mathrm{M}_\odot$ range \citep{2023MNRAS.526.4978S}. Furthermore,  the directional sensitivity of the scattering coefficients may amplify this effect of weaker suppression, since the anisotropic nature of jets leads to a less uniform redistribution of gas compared to the isotropic thermal AGN mode.

Second-order scattering coefficients reveal that feedback-induced suppression of clustering is most pronounced on small scales. At $j \geq 3$, the clustering suppression becomes largely independent of $j_2$: $\mathcal{T}^{j_1, j_2}_2 \approx \mathcal{T}^{j_1}_1$. This differs from the thermal model, which more strongly reduces correlations between structures of similar size. Moreover, the jet feedback shows a consistently stronger suppression on larger scales compared to smaller ones, which contrasts with the fiducial scenario. This feature of the jet model implies that clustering suppression is influenced not only by the gas fraction but also by the extent of spatial gas redistribution, in line with the findings from \cite{2020MNRAS.492.2285D}.

However, when the jet model becomes more aggressive, $\mathrm{Jet}\_{\rm fgas} - 4\sigma$, its suppression exceeds that of the thermal models across all scales, with transfer function values falling below even the extreme thermal AGN feedback case (fgas$-8\sigma$) at $j = 0-1$. For $j_1 \geq 2$, differences between the models disappear. This stronger clustering suppression in $\mathrm{Jet}\_{\rm fgas} - 4\sigma$ may be related to the directional sensitivity of the scattering transform.
Importantly, if these small scales are unresolved, two physically different models, $\mathrm{Jet}\_{\rm fgas} - 4\sigma$ and kinetic fgas$-8\sigma$, become effectively indistinguishable, emphasising the importance of high-resolution data for discriminating feedback mechanisms.

\subsubsection{Noise and smoothing}

\begin{figure*}
    \centering
    \includegraphics[width=\textwidth]{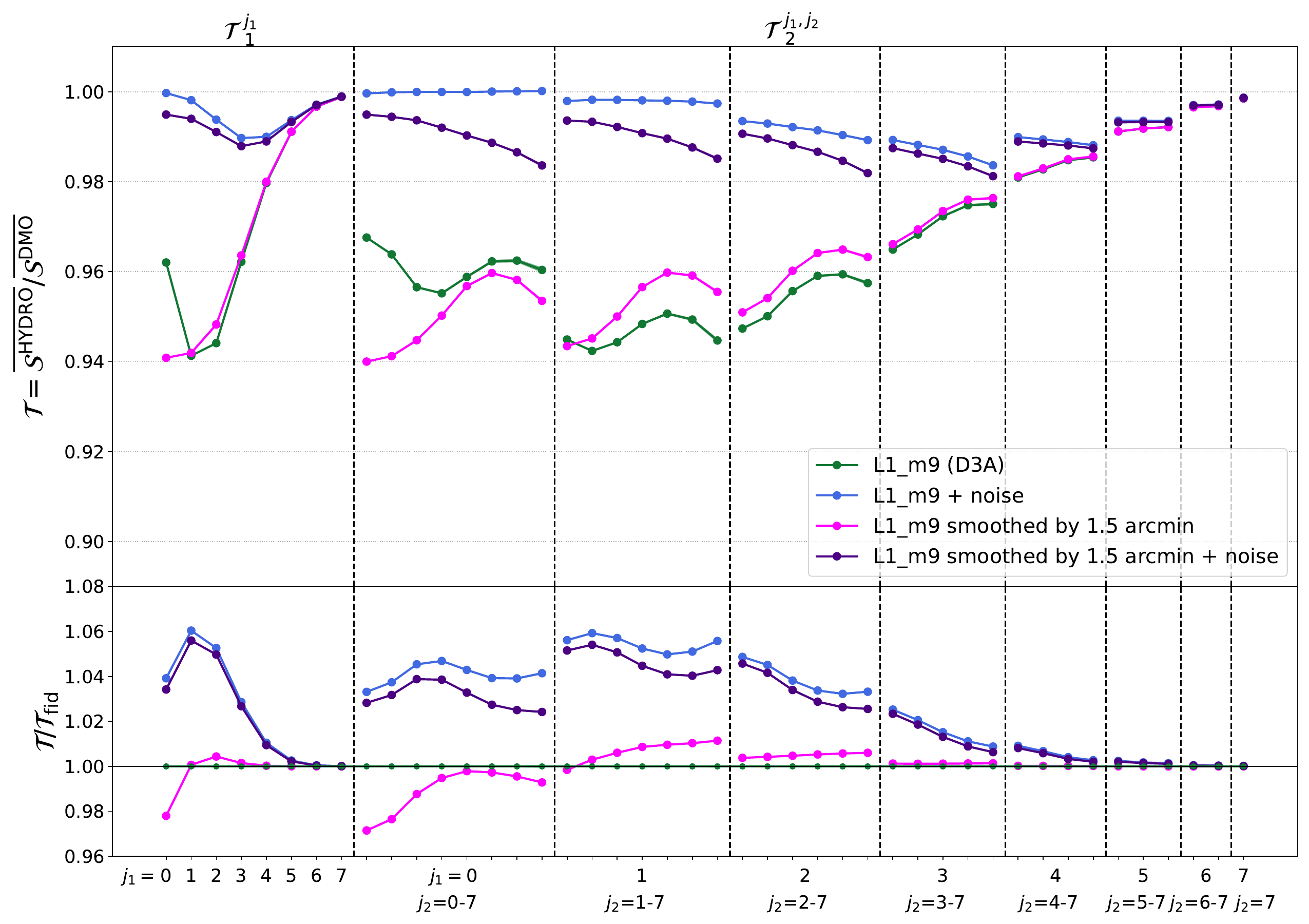}
    \caption{Scattering coefficients $\mathcal{S}$ as a function of dyadic scale for the D3A cosmology with fiducial feedback mode shown in four cases: noiseless and unsmoothed (L1\_m9), noise only, smoothed only, and both noise and smoothing applied. Smoothing suppresses small-scale power by averaging over local fluctuations, thereby reducing the amplitude of coefficients at low $j$ while leaving large-scale modes largely unaffected. Shape noise dominates on small scales ($j \leq 2$), pushing the scattering coefficients to values determined by noise and obscuring physical structure. }
    \label{fig:tf_fiducial_noise_smoothing}
\end{figure*}

The results presented above assume idealised conditions without instrumental noise or smoothing. To imitate more realistic observational settings, we introduced galaxy shape noise at the pixel level by sampling from a Gaussian distribution $\mathcal{N}\big\{\mu, \sigma\big\}$ with zero mean ($\mu = 0$) and a standard deviation $\sigma$~\citep{1993ApJ...404..441K, 2021MNRAS.506.3406L}:
\begin{equation}
    \sigma = \frac{\sigma_\epsilon}{\sqrt{2 n_\mathrm{gal}A_\mathrm{pix}}},
\end{equation}
where $\sigma_\epsilon = 0.26$ is the intrinsic ellipticity dispersion, $n_\mathrm{gal} = 30~\mathrm{arcmin}^{-2}$ is the galaxy number density, consistent with expected observational characteristics of \textit{Euclid}~\citep{2011arXiv1110.3193L}, and $A_\mathrm{pix}$ is the pixel area. The noise is added to each pixel of the unsmoothed images, both for DMO and hydrodynamical maps.

To study the impact of angular smoothing, we further convolve the original and noisy maps with a Gaussian filter of 1.5 arcmin full width at half maximum (FWHM). This smoothing scale is equivalent to approximately 4 pixels, slightly larger than the 1 arcmin used in  \cite{PhysRevD.91.063507}, which is motivated by a trade-off between preserving small-scale structures and alleviating noise. In our case, the choice of a broader kernel is made deliberately to explore the effect of smoothing on the transfer function. 

Figure~\ref{fig:tf_fiducial_noise_smoothing} shows the effect of smoothing and shape noise on the baryonic transfer function $\mathcal{T}^{j_1, j_2}_2$, computed for the fiducial D3A cosmology with fiducial AGN feedback. Smoothing acts by averaging over local fluctuations, effectively suppressing high-frequency features in the maps. This results in a general decrease in the amplitude of scattering coefficients, especially for the smallest scale, $j_1 = 0$, where the wavelets are sensitive to the finest structures. Interestingly, for scales just below the smoothing kernel size ($j_1=1-2$), the amplitude of $\mathcal{T}$ may increase slightly, while larger scales remain unaffected. This behaviour occurs because, in hydrodynamical simulations, small-scale structure is already suppressed by baryonic feedback, making the scattering coefficients more sensitive to additional smoothing. On the smallest scales, where the structures are enhanced, smoothing washes out these features. On small to intermediate scales, where baryonic feedback suppresses power, smoothing mitigates this effect by reducing the contrast between DMO and hydro simulations. As a result, smoothing wipes out the observable imprint of baryonic feedback across a broad range of scales. For example, in the presence of noise, the fgas$-8\sigma$ and $\mathrm{Jet\_fgas}4-\sigma$ models become practically indistinguishable, which is not shown in the plots.  These statements hold for both first- and second-order coefficients.

The inclusion of shape noise alters the transfer function more drastically. On small scales, the noise dominates over the physical signal, pushing $\mathcal{T}$ towards unity and reducing the suppression to below $\sim2\%$, even at intermediate scales. At $j \leq 2$, the wavelets primarily measure the noise itself, producing nearly identical scattering coefficients in DMO and hydro simulations and making the transfer function uninformative.  Beyond $j=4$, the ratios of scattering coefficients between cosmological models within the same simulation type (hydro or DMO) are primarily determined by the filter size, rather than the physical differences between the models. This demonstrates that shape noise systematically lowers the sensitivity of the scattering transform to baryonic feedback across all scales. 

Overall, while smoothing primarily removes small-scale cosmological information, galaxy shape noise has a more profound impact. For the first-order scattering coefficients, information is retained at $j=0-4$, where baryon modelling is the dominant systematic. However, at $j \geq 5$, shape noise and baryon modelling impact the measurement with roughly the same strength, and the transfer function may even become irrelevant once shape noise dominates over baryonic effects. The usefulness of the transfer function for Euclid-like surveys depends strongly on data quality, and our results cannot provide a definitive answer to the performance of the transfer function for parameter inference in real surveys. On the other hand, $\mathcal T =1$ is desirable for probing primary physics. These results highlight the importance of high-resolution data and noise mitigation strategies for maximising constraining power of  WL surveys. 

\section{Summary}

\label{sec:summary}

In this work, we have examined the impact of baryonic feedback on the scattering transform of weak lensing convergence maps with \textit{Euclid}-like source redshift distribution from the FLAMINGO suite of hydrodynamical and DMO simulations. Our analysis is focused on the transfer function, which is the ratio of ST coefficients in hydrodynamical simulations to those in their DMO counterparts, and its sensitivity to the baryonic effects. 

For our analysis, we divided each full-sky map into $3183$ sky patches of $3.6 \times 3.6$ deg$^2$ and computed reduced scattering coefficients up to second order for each patch using the pipeline from \citet{2020MNRAS.499.5902C}. Since all the simulations have the same initial conditions and the observer position is the same for the studied convergence maps, we could make a comparison, avoiding cosmic variance. While the scattering coefficients themselves show up to $\sim5 -10\%$ sensitivity to cosmology and baryonic physics, we showed that the corresponding transfer functions between hydrodynamical and DMO runs are largely insensitive to cosmology but remain strongly responsive to baryonic feedback.

Our key findings are:

\begin{itemize}
    \item Scattering coefficients react strongly to the changes in $S_8$ (Figure~\ref{fig:sc_all_cosmologies}), whose lower values lead to suppressed amplitudes across all scales. This effect is approximately scale-independent, indicating that cosmological information is incorporated through a global rescaling of the coefficients.

    \item Baryonic feedback causes scale-dependent suppression of the scattering coefficient, which is the most prominent at small and intermediate scales ($j=1-3$), reaching up to $4\%$ for the extreme feedback models (Figure~\ref{fig:sc_all_feedbacks}). The magnitude of this effect is comparable to that of cosmological variations and must be properly accounted for in precision inference.

    \item The transfer function $ \mathcal T =\frac{ \overline{\mathcal{S}^{\mathrm{HYDRO}}} }{ \overline{ \mathcal{S}^{\mathrm{DMO}}}}$ is stable across pacthes (Figure~\ref{fig:TF_hist_combined}), following a Gaussian distribution with low mean-to-variance ratio of $\lesssim 0.4\%$. 

    \item Transfer functions are largely independent of changes in cosmology, with mean differences relative to the fiducial cosmology remaining below $ 0.3\%$ for the fixed feedback model (Figure~\ref{fig:TF_hist_cosm}).  This conclusion holds across all scales and cosmological models (Figure~\ref{fig:full_tf_cosmol}), demonstrating that although cosmology affects the absolute amplitude of the scattering coefficients, the relative ratio $\cal T$ remains effectively invariant. In contrast, transfer functions show clear sensitivity to baryonic effects, as indicated by mean variations exceeding  $ 4\%$ (Figure~\ref{fig:TF_hist_feedback}). These significant differences are observed across all scales (Figure~\ref{fig:full_tf_feedback}).

    \item We explored nine baryonic feedback models that systematically vary the galaxy stellar mass function, cluster gas fraction, and AGN feedback implementation (Figure~\ref{fig:full_tf_feedback}). Thermal AGN  feedback produces strong suppression of the scattering coefficients on small and intermediate scales ($j=1-5$, or $\sim 2 - 10$ cMpc), while jet AGN feedback results in more extended but weaker suppression. The response of the scattering coefficients to changes in $f_{\mathrm{gas}}$ is monotonic but non-linear, and modifications of the stellar mass function amplify it further, especially on small scales. Moreover, certain feedback variations ($f_{\mathrm{gas}} - 8 \sigma$ and $\mathrm{Jet\_fgas} - 4\sigma$) are distinguishable only on the smallest scales, emphasising the need for high-resolution measurements to constrain baryonic feedback. 

    \item We showed that galaxy shape noise strongly degrades the sensitivity of the scattering transform to baryonic feedback, washing out small-scale structure and driving the transfer function toward unity (Figure~\ref{fig:tf_fiducial_noise_smoothing}). This effect cannot be completely counterbalanced even by 1.5 arcmin smoothing, retaining some physical suppression at $j=2-5$. It once again indicates the importance of high-resolution data and noise mitigation for weak lensing surveys. 
    
\end{itemize}

These results are consistent with previous findings by \citet{2024PhRvD.110j3539G}, who computed baryonic impact on the weak lensing scattering transform using the IllustrisTNG and BAHAMAS hydrodynamical simulations, and by \citet{2024MNRAS.529.2309B}, who studied baryonic effects on peak counts using the same FLAMINGO data as used here. While finalising this work, we also became aware of an independent study by \citet{2025arXiv250507949Z}, who computed a similar transfer function using FLAMINGO simulations in a DES-like survey configuration. Their findings on the ratios of scattering coefficients between hydrodynamical and DMO simulations, computed across varying cosmologies and baryonic feedback models, are consistent with ours and further support our conclusions.

Together, these results demonstrate that the scattering transform is a robust and physically interpretable statistic for capturing the imprint of baryonic processes in weak lensing convergence maps. The stability, scale-dependent structure, and minimal sensitivity to cosmology of the transfer function $\mathcal{T}$ make it a valuable tool for modelling baryonic effects in cosmological analyses. In particular, its predictive nature motivates its use as a correction factor in likelihood-free inference, emulation, or Bayesian model averaging analysis, helping to mitigate baryonic effects and recover cosmological information on small scales.

While this study is focused on redshift-integrated convergence maps, the redshift dependence of baryonic effects can be further explored with tomographic analyses. This will be addressed in future work, where we intend to extend the transfer function formalism to tomographic bins and examine its redshift evolution.



\section*{Data Availability}

The baryonic transfer function results presented in this paper, including analogues for all wavelet types discussed in Appendix~\ref{appendix:wavelets} as well as for additional patch sizes ($1.8 \times 1.8$, $5 \times 5$, and $7.5 \times 7.5$ deg$^2$), are available in the GitHub repository: \href{https://github.com/Mariia-Marinichenko/BF-ST-transfer-functions}{https://github.com/Mariia-Marinichenko/BF-ST-transfer-functions}. Results for other configurations can be provided upon reasonable request to the authors.




\bibliographystyle{mnras}
\bibliography{references,Bib} 




\appendix
\section{Wavelet families}
\label{appendix:wavelets}
The key component of the scattering transform is the wavelet filter bank. Wavelets used in the scattering transform must satisfy the following admissibility criteria to ensure a stable and informative multiscale representation. They must be complex-valued to provide approximate shift invariance and directional selectivity. For energy conservation and effective information extraction, the wavelets must uniformly cover the Fourier space, excluding null frequency. Furthermore, their bandwidths must scale proportionally with their central frequencies, maintaining constant relative frequency resolution and ensuring robustness to small deformations. Together, these conditions guarantee both the completeness and stability of the scattering representation.

The wavelet family is generated by dyadic dilations (with the factor of 2) and discrete rotations of a mother wavelet $\psi(\boldsymbol{x})$.  For a given scale index $j \in [0,\ \mathrm{int}(\log_2 M)]$, where M is the image resolution in pixels, and an orientation index $l$, the wavelet $\psi^{j,l}(\boldsymbol{x})$ is defined as:
 \begin{equation}
 \psi^{j,l}(\boldsymbol{x}) = 2^{-2j} \cdot \psi\left( 2^{-j}r_l^{-1} \boldsymbol{x}\right),
 \end{equation}
 where $r_l$ denotes a rotation by angle $\pi\: l/L$ with $L$ being the total number of discrete orientations and $l$ varying from $1$ to $L$.
 
In the Fourier domain, this corresponds to:
 \begin{equation}
 \hat{\psi}^{j,l}(\boldsymbol{k}) = \hat{\psi}\left( 2^{j}r_l^{-1} \boldsymbol{k}\right).
 \end{equation}
The mother wavelet is centered at $\boldsymbol k_0$ frequency with bandwidth equal to one, menaning that $\hat{\psi}^{j,l}(\boldsymbol{k})$ is centered at $2^{-j}\boldsymbol{k}_0$, with spectral support scaling as $2^{-j}$.

In our analysis, we use Morlet wavelets (also known as Gabor wavelets), a widely used choice in prior scattering transform applications:
\begin{equation}
        \psi(\boldsymbol{x}) =\frac{1}{\sqrt{|\boldsymbol{\Sigma}|}} e^{-\frac{1}{2}\boldsymbol{x}^T \boldsymbol{\Sigma}^{-1} \boldsymbol{x} }\big(e^{i \boldsymbol{k_0} \cdot \boldsymbol{x}} -\beta \big),
    \end{equation}
with $\beta = e^{-\frac{1}{2}\boldsymbol{k_0}^T \boldsymbol{\Sigma}^{-1} \boldsymbol{k_0} }$, a factor that cancels the null frequency. This wavelet is a combination of a plane wave of wavenumber $k_0$ modulated by a Gaussian envelope with the covariance matrix $\boldsymbol{\Sigma}$. The matrix is designed to have eigenvalues $\sigma^2$ and $\sigma^2/s^2$, with $\sigma^2$ aligned along $\boldsymbol{k_0}$. The parameter $s$ controls the ellipticity of the wavelet in Fourier space.

The Fourier representation of the wavelet $ \psi(\boldsymbol{x})$ takes the form:
\begin{equation}
         \hat{\psi}(\boldsymbol{k}) = e^{-\frac{1}{2}\boldsymbol{(k-k_0)}^T \boldsymbol{\Sigma}^{-1} \boldsymbol{(k-k_0)} } -\beta\: e^{-\frac{1}{2}\boldsymbol{k}^T \boldsymbol{\Sigma}^{-1} \boldsymbol{k} } .
\end{equation}

For the 2D case, the parameters are chosen as:
    \begin{equation}
        \sigma = 0.8 \times 2^j, \quad k_0 \equiv|\boldsymbol{k_0}| = \frac{3\pi}{4 \times 2^j}, \quad\theta_0=\frac{\pi l}{L}, \quad s = \frac{4}{L},
    \end{equation}
where $\sigma$ is given in pixels, and $\theta_0$ is angle between $\boldsymbol{k}$ and $\boldsymbol{k_0}$. These parameter choices follow the default settings of the \texttt{kymatio} scattering transform package\footnote{\hyperlink{https://www.kymat.io/userguide.html}{https://www.kymat.io/userguide.html}} and are designed to ensure that the number of oscillations within each wavelet remains approximately constant across scales: $k_0\sigma\approx2$.

Although we show only the results for Morlet wavelets, the Sihao's scattering transform package has more directional wavelets, results for which are available at the GitHub page: \href{https://github.com/Mariia-Marinichenko/BF-ST-transfer-functions}{https://github.com/Mariia-Marinichenko/BF-ST-transfer-functions}. Below, we list these wavelets:

\begin{itemize}

    \item Bump steerable wavelet:
    \begin{equation}
        \hat{\psi}(\boldsymbol{k}) = \frac{2^{L/2-1}(L/2-1)!}{1.29 \sqrt{(L/2)(L-2)!}} e^{-\frac{(k - k_0)^2}{k_0^2 - (k - k_0)^2} } \cos^{L/2 - 1}(\theta - \theta_0),
    \end{equation}
defined for $k \in [0, 2k_0]$ and $\theta \in [\theta_0 - \frac{\pi}{2}, \theta_0 + \frac{\pi}{2}]$, and zero elsewhere.

This steerable band-pass filter is particularly effective for edge detection and orientation-sensitive tasks \citep[][]{2018arXiv181012136M}.
 
    \item Gaussian steerable (circular harmonic) wavelet:
    \begin{equation}
        \psi(\boldsymbol{k}) = \frac{2^{L/2-1}(L/2-1)!}{1.29 \sqrt{(L/2)(L-2)!}} \left( \frac{2k}{k_0} \right)^2 e^{ -\frac{\big(1.4\:k\big)^2}{2\: k_0^2}} \cos^{L/2+1}(\theta - \theta_0),
    \end{equation}
    with $\theta \in [\theta_0 - \frac{\pi}{2}, \theta_0 + \frac{\pi}{2}]$.
This wavelet highlights blob-like structures with a characteristic radius $k_0 \sqrt{2}/1.4$. By combining isotropic and directional properties, it is well-suited for detecting rounded features and wedges~\citep[][]{2019arXiv190712165K}.

    \item Shannon wavelet:
    \begin{equation}
        \psi(\boldsymbol{k}) = H(k > k_{\min}) \, H(k < k_{\max}) \,H(\theta - \theta_0 + \frac{\pi}{2}) \, H(\theta_0 - \theta+\frac{\pi}{2}),
    \end{equation}
    where $H(x)$ is the Heaviside step function.

This rectangular band-pass filter sharply selects both frequency and angular bands, which is useful for isolating targeted features while avoiding spectral leakage~\citep[][]{mallat1999wavelet}.

    \item Gaussian harmonic:
    \begin{equation}
        \psi(\boldsymbol{k}) = \left( \frac{2k}{k_0} \right)^2 e^{ -\frac{k^2}{2(k_0/1.4)^2} } e^{i l \theta},
    \end{equation}
    where $l$ is the harmonic index.
This wavelet incorporates angular modulation through $e^{i l \theta}$, enabling the detection of rotational symmetries and phase-dependent patterns in the signal.

\end{itemize}

\begin{figure}
    \centering
    \includegraphics[width=\linewidth]{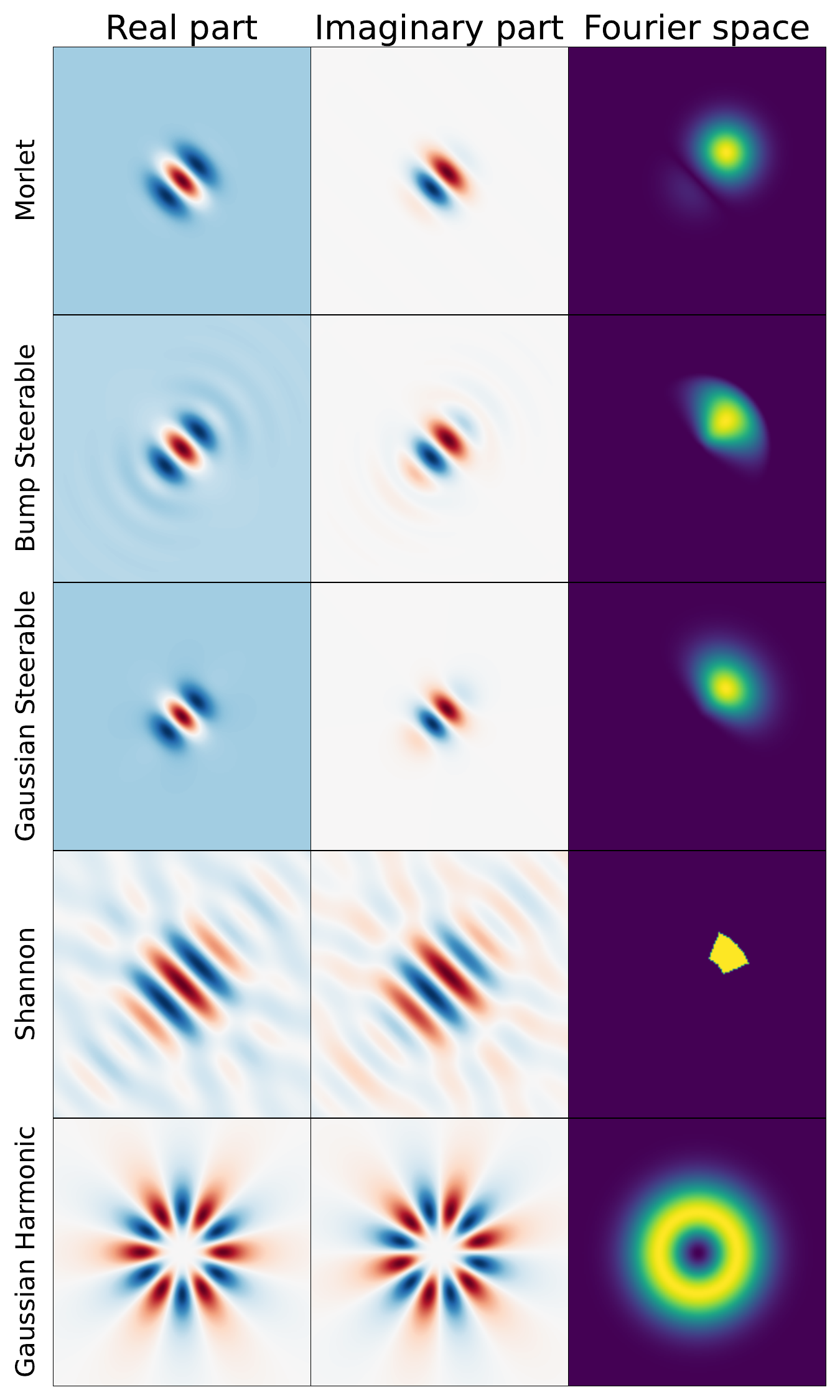}
    \caption{Real part (left), imaginary part (middle), and Fourier modulus (right) of five wavelet types implemented in the scattering transform package. From top to bottom: Morlet, Bump Steerable, Gaussian Steerable, Shannon, and Gaussian Harmonic. All wavelets are shown for the same central frequency and orientation to enable direct comparison.}
    \label{fig:filter_plots}
\end{figure}

All five wavelet types are illustrated in Figure~\ref{fig:filter_plots}, each shown at the same central frequency and orientation for direct visual comparison. For each wavelet, we display the real and imaginary parts in real space (left and middle columns), along with the corresponding amplitude in Fourier space (right column). This layout highlights differences in spatial localisation, phase structure, and spectral extent across various wavelet types.
\section{$S_8$ dependence of scattering coefficients}
\label{appendix:S_8_dependence}

The FLAMINGO simulation suite includes five different cosmological models that vary in three main parameters: the total matter density fraction $\Omega_\mathrm{m}$, present-day linear rms fluctuation in 8 Mpc h-1 volume $\sigma_8$, and the sum of neutrino masses $\sum m_\nu$. All other cosmological parameters either remain fixed or are derived from these three. The simulations that include baryons (HYDRO) have an additional parameter -- the baryon density fraction, $\Omega_b$. However, the baryon fraction, defined as $\Omega_\mathrm{b}/\Omega_\mathrm{m}$, remains nearly constant across all cosmologies ($\approx 0.156$–$0.159$), effectively reducing the number of independent parameters to three.

Furthermore, the parameters $\Omega_\mathrm{m}$ and $\sigma_8$ can be combined into a single parameter, $S_8 = \sigma_8 \sqrt{\Omega_\mathrm{m}/0.3}$, which captures the degeneracy direction along which the matter and, hence, weak lensing convergence power spectra are most sensitive. This reduces the effective cosmological parameter space to two dimensions. The neutrino mass $\sum m_\nu$ is varied in such a way that the remaining parameters remain close to the Planck best-fit values, and its influence on the matter power spectrum is minimal~\citep[Figure 21 in][]{2023MNRAS.526.4978S}. As a result, the weak lensing convergence maps, and the scattering coefficients derived from them, depend primarily on $S_8$. 

\begin{figure}
    \centering
    \includegraphics[width=\columnwidth]{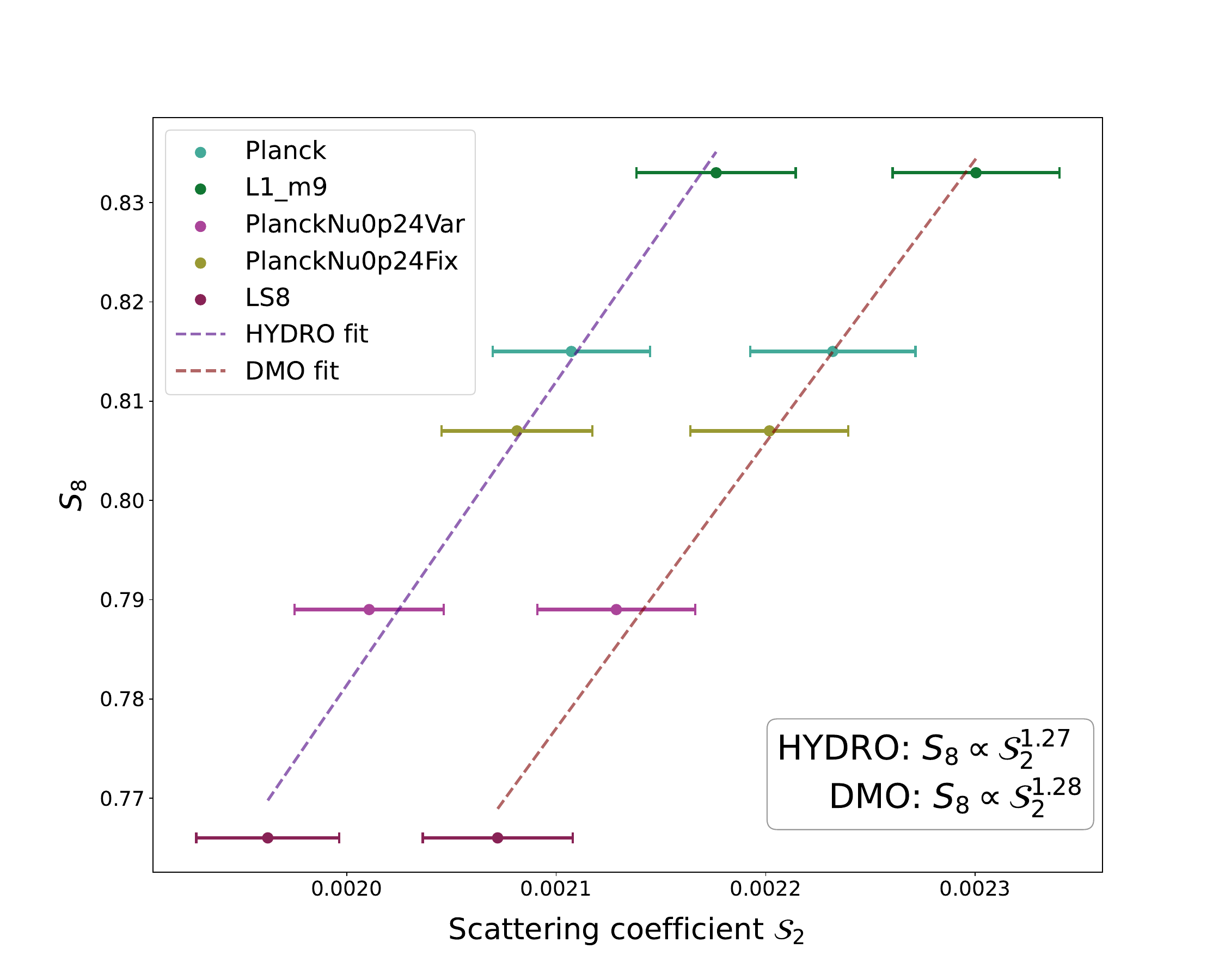}
    \vspace{0.5em}
    \includegraphics[width=\columnwidth]{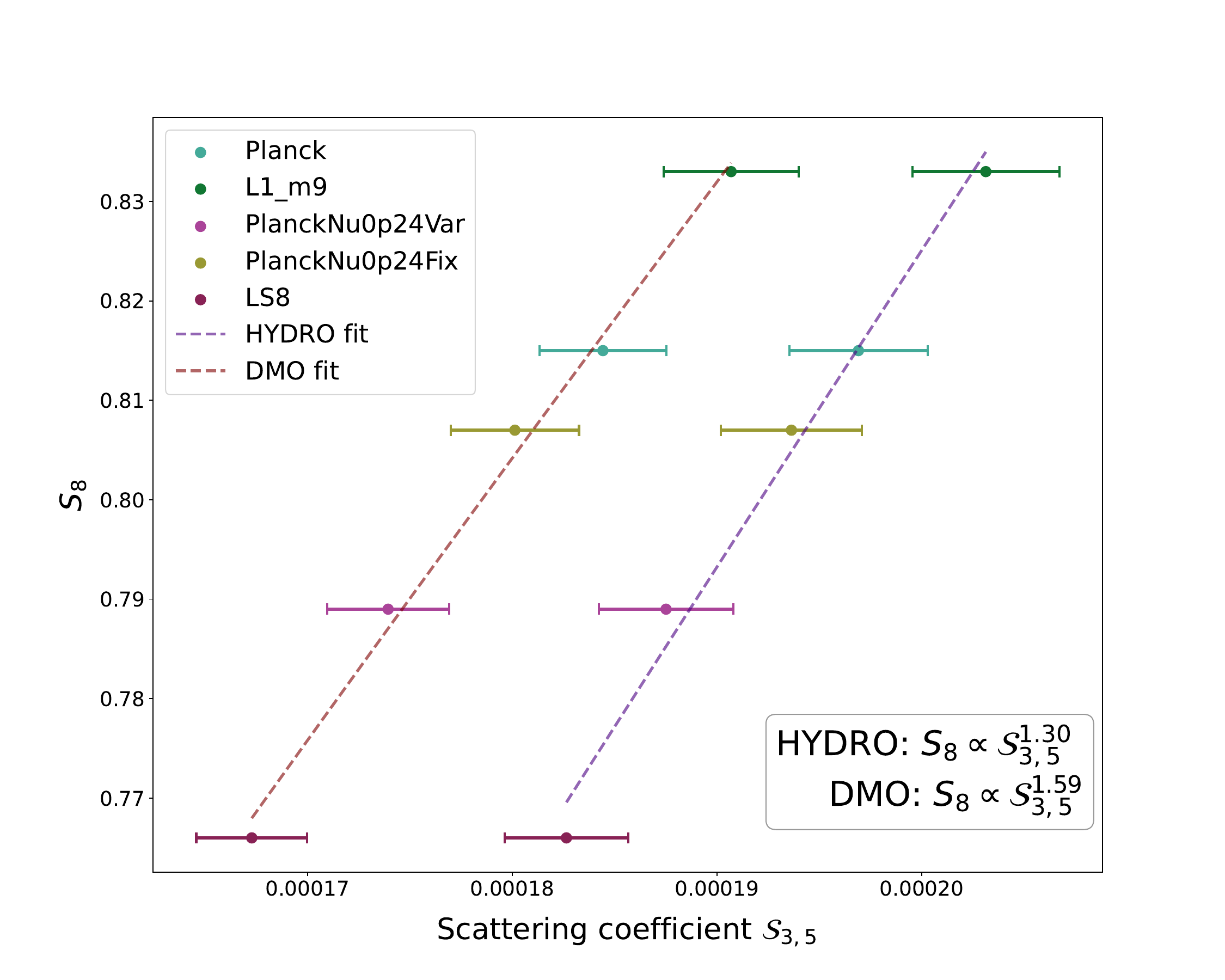}
    \caption{Dependence of selected scattering coefficients on the cosmological parameter $S_8 = \sigma_8 \sqrt{\Omega_\mathrm{m} / 0.3}$.  Each point corresponds to the patch and orientation averaged coefficient measured from either DMO or HYDRO convergence maps. Top:  First-order coefficient $\mathcal{S}_2$.
    Bottom: Second-order coefficient $\mathcal{S}_{3,5}$.}
    \label{fig:S_8_ratio}
\end{figure}

We computed the scattering coefficients across cosmologies and found that they demonstrate a clear power-law dependence on $S_8$, with slopes typically in the range $\sim 1.0$–$1.6$, staying consistent between DMO and HYDRO simulations. The second-order coefficients display a slightly larger variation in slope than the first-order ones, indicating that cross-scale correlations are a bit more sensitive to baryonic processes. Figure~\ref{fig:S_8_ratio} illustrates this relationship for two representative coefficients, $\mathcal{S}_2$ and $\mathcal{S}_{3,5}$.

\bsp	
\label{lastpage}
\end{document}